\documentclass[twoside,11pt,letterpaper]{article}
\usepackage{jair}
\usepackage[natbibapa]{apacite}
\renewcommand{\APACrefbtitle}[2]{\Bem{#1}} %Fixes apacite error

\newcommand{\shortcitep}[1]{(\citeyear{#1})}

\usepackage{graphicx}
\usepackage[usenames,dvipsnames]{xcolor}
\usepackage{enumerate}

\usepackage{booktabs}

\usepackage{url}
\usepackage{amsmath}
\usepackage{mathtools}
\usepackage{xspace}
 \newcommand{\etal}{{et~al.\@}\xspace}

\usepackage[inline]{enumitem}
\setlist{nosep}%,leftmargin=0.5em

\DeclareRobustCommand{\nAmE}[3]{#2}

\usepackage{xcolor}
\definecolor{labelcolor}{cmyk}{0.22,0.10,0.10,0.10}
\definecolor{listbackgroundcolor}{cmyk}{0.10,0.10,0.05,0.05}
\definecolor{listbackgroundcolorlight}{rgb}{0.91,0.92,0.94}
\definecolor{colorEntityBack}{rgb}{0.01,0.01,0.4}
\definecolor{colorPolicyBack}{rgb}{0.91,0.94,0.94}
\definecolor{colorApproachBack}{rgb}{0.93,0.93,0.97}

\definecolor{colorRepresentationBack}{rgb}{0.81,0.81,0.86}%
\definecolor{colorGreen}{rgb}{0.07, 0.43, 0.07}%

\newtheorem{example}{Use Case}
\newcommand{\pr}{Use Case}

\newtheorem{constraint}{Constraint}

\newtheorem{assumption}{Assumption}

\newtheorem{principle}{Principle}

\usepackage{listings}
\newcommand{\ar}[1]{\ensuremath{\raisebox{-2pt}{$\xrightarrow{\text{#1}}$}}}

\usepackage{tikz}
\usetikzlibrary{calc,backgrounds,arrows,shapes}

\usetikzlibrary{positioning}
\usetikzlibrary{arrows,shapes,automata,petri}
\usetikzlibrary{matrix}
\usetikzlibrary{fit}
\usetikzlibrary{decorations}
\usetikzlibrary{decorations.markings}
\usetikzlibrary{calc}
\usetikzlibrary{graphs}
\usetikzlibrary{quotes}
\usetikzlibrary{arrows.meta}

\tikzstyle{thrust}=[circle,font=\footnotesize\sffamily,fill=colorRepresentationBack,text=colorEntityBack,solid,draw=red, thick,minimum width=12]%

\tikzstyle{message}=[->,-{Classical TikZ Rightarrow[length=1.5mm]},>=stealth']
\usepackage{subcaption}
\usepackage{mathtools}

\usepackage{amssymb}
\newcommand{\ul}{\ulcorner}
\newcommand{\ur}{\urcorner}
\newcommand{\inn}{\ensuremath{\ul\mathsf{in}\ur}\xspace}
\newcommand{\out}{\ensuremath{\ul\mathsf{out}\ur}\xspace}
\newcommand{\nil}{\ensuremath{\ul\mathsf{nil}\ur}\xspace}

\newcommand{\mo}{\mapsto}

\newcounter{mpscount}

\newcommand{\fbf}{\textbf}
\newcommand{\fsl}{\textsl}

\newcommand{\fsf}[1]{\textsf{#1}}
\newcommand{\msf}{\mathsf}

\newcommand{\role}{\ensuremath\msf{role}}
\newcommand{\param}{\ensuremath\msf{parameter}}

\newcommand{\mname}[1]{\fsl{#1}}
\newcommand{\pname}[1]{\fsl{#1}}
\newcommand{\rname}[1]{\textsc{#1}}
\newcommand{\val}[1]{\texttt{#1}}
\newcommand{\paraname}[1]{\fsf{#1}}

\newcommand{\speciallabelsize}{\normalsize\rm}

\usepackage{marginnote}
\newcommand{\review}[1]{}

\newcommand{\responsePtr}[1]{}

\newcommand{\commit}{\ensuremath\msf{commitment}}
\newcommand{\toward}{\ensuremath\msf{to}}

\newcommand{\create}{\ensuremath\msf{create}}

\newcommand{\detach}{\ensuremath\msf{detach}}
\newcommand{\discharge}{\ensuremath\msf{discharge}}

\ShortHeadings{Evaluation of Communication Protocol Languages for Engineering MAS}
{Chopra, Christie, \& Singh}
\firstpageno{1}

\begin{document}
%\blindtext[5]

\title{An Evaluation of Communication Protocol Languages for Engineering Multiagent Systems}

\author{\name Amit K.~Chopra \email amit.chopra@lancaster.ac.uk \\
	\addr Lancaster University\\Lancaster, LA1 4WA, UK
        \AND
	\name Samuel H.~Christie V \email schrist@ncsu.edu \\
	\addr North Carolina State University\\Raleigh, NC 27695, USA\\
        \addr Lancaster University\\Lancaster, LA1 4WA, UK
        \AND
	\name Munindar P.~Singh \email singh@ncsu.edu \\
	\addr North Carolina State University\\Raleigh, NC 27695,
	USA
}

\maketitle

\begin{abstract}
  \review{B:abstract} Communication protocols are central to
  engineering decentralized multiagent systems.  Modern protocol
  languages are typically formal and address aspects of
  decentralization, such as asynchrony.  However, modern languages
  differ in important ways in their basic abstractions and operational
  assumptions.  This diversity makes a comparative evaluation of
  protocol languages a challenging task.

  We contribute a rich evaluation of modern protocol languages based
  on diverse approaches.  Among the selected languages, Scribble is
  based on session types; Trace-C and Trace-F on trace expressions;
  HAPN on hierarchical state machines, and BSPL on information
  causality.  Our contribution is four-fold.  One, we contribute
  important criteria for evaluating protocol languages.  Two, for each
  criterion, we compare the languages on the basis of whether they are
  able to specify elementary protocols that go to the heart of the
  criterion.  Three, for each language, we map our findings to a
  canonical architecture style for multiagent systems, highlighting
  where the languages depart from the architecture.  Four, we identify
  a few design principles for protocol languages as guidance for
  future research.
\end{abstract}

%%   We find that BSPL, which models protocols declaratively in terms
%%   of information, makes weaks operational assumptions than the
%%   other languages and is able to capture scenarios that they
%%   cannot.

%% All languages except BSPL perform poorly in our evaluation.

\section{Introduction}
\label{sec:intro}

We understand a multiagent system (MAS) as a decentralized system of
autonomous agents, each of whom represents a real-world entity such as
a person or an organization.  In particular, a MAS is not a separate
computational entity but is realized purely through its member agents.
When a MAS models a collective such as an
institution,
\review{B:institution}
the institution can be viewed as an
entity that is itself a member of the MAS \citep{TIST-13-Governance}.

Agents in a MAS coordinate their computations while retaining loose
coupling in their construction and decision making.
\review{B:explicit}
To accommodate such a conception of a MAS, it is
crucial that (1) agents coordinate their computations via arms-length
communication, that is, via \emph{asynchronous messaging}, and (2) the
coordination requirements be clearly specified and support a
programming model that facilitates the construction of agents.

The foregoing twin requirements have motivated an extensive study of
languages for specifying \emph{communication protocols}.
\review{C:protocol} Broadly, a communication protocol specifies the
coordination in a MAS by specifying two or more interacting roles, the
messages (that is, message schemas) exchanged by these roles, and the
conditions under which agents playing those roles may send (instances
of) the various messages to one another.  Further, a protocol yields a
programming interface (skeleton) for every agent such that if an agent
implements its interface, then the agent is \emph{compliant}.
Compliance with the stated protocol is the primary \emph{correctness}
criterion for individual agents: If all agents complied, then the
desired coordination between them would obtain.

For concreteness, \pr~\ref{pr:purchase} describes a protocol
informally.

\begin{example}[Purchase]
  \label{pr:purchase}
  A buyer requests an item from a seller, who responds with an offer.
  The buyer may accept or reject the offer.  If the buyer accepts the
  seller's offer, the seller delivers the specified item to the buyer,
  following which the buyer sends the specified payment to the seller.
\end{example}

For \pr~\ref{pr:purchase}, the roles would be \rname{buyer} and
\rname{seller}.  \review{B:message}The messages would be
\mname{Request}, \mname{Offer}, \mname{Reject}, \mname{Accept},
\mname{Payment}, and \mname{Deliver}.  And, \rname{buyer} may send
\mname{Request}, \mname{Accept}, \mname{Reject}, and \mname{Payment},
and \rname{seller} may send \mname{Offer} and \mname{Deliver}.  Each
message contains the information relevant to that message (presumably
capturing what that message connotes in the protocol).  For example,
\mname{Request} contains the \paraname{item} and \mname{Offer}
contains the \paraname{price}.  We identify additional constraints
from \pr~\ref{pr:purchase}: \mname{Offer} concerns the same
\paraname{item} as specified in \mname{Request}; and, \mname{Payment}
specifies the same amount as the \paraname{price} in \mname{Offer}.

The notion of protocol is naturally a foundational one for multiagent
systems.  Following \citet{Hewitt91}, \citet{Gasser-91} identifies
protocols as one of the central challenges for MAS.  Protocols are
central to work on agent communication languages
\citep{FIPA-ACL-2002,vieira:speech-acts:2007}; on institutions, e.g.,
\citep{dinverno:institutions:2012}; and on agent-oriented software
engineering (AOSE) methodologies such as Gaia \citep{Zambonelli+03},
Tropos \citep{Bresciani-Tropos-JAAMAS-04}, and Prometheus
\citep{winikoff:prometheus:2005}.

The centrality of protocols has spurred work on protocol specification
languages and approaches.  \review{B:AUML-b} Agent UML (AUML)
\citep{Odell+2001}, an early graphical notation for specifying
protocols that extends UML, is applied in Tropos and Prometheus and by
FIPA for specifying interaction protocols \citep{FIPA-IP}.
\review{C:FIPA}
Inspired
from problems in telecommunications networks and UML, Message Sequence
Charts \citep{ITU-MSC-04} was another early standardization of a
protocol notation.

% Well-known early approaches for protocol specification, e.g.,
% \citep{ITU-MSC-04,Vitteau-protocols-2004} did not typically address
% asynchronous protocol enactments.

% \marginnote{\customlabel{foobar}{22}}

\subsection*{Problem}

From modest beginnings in informal notations based on UML, work on
protocol languages has grown to encompass a number of
sophisticated---and in some cases, complicated---formal approaches
that boast diverse basic abstractions and operational assumptions.
For example, the abstractions employed include state machines
\citep{Baldoni+06,Winikoff+18:HAPN}, logic-based constraints
\citep{baldoni:regulative:2013}, action descriptions
\citep{AAAI-Enact-08}, trace expressions
\citep{Castagna+12:sessions,ferrando:enactability:2019}, session types
\citep{Yoshida+13:Scribble}, and information constraints
\citep{AAMAS-BSPL-11}.  Operational assumptions range from synchronous
communication \citep{Winikoff+18:HAPN}, to asynchronous but pairwise
FIFO communication \citep{Castagna+12:sessions,Yoshida+13:Scribble},
to unordered asynchronous communication \citep{AAMAS-BSPL-11}.

The rich diversity of languages for specifying protocols raises an
important question: how may we compare and evaluate them?  Today, we
lack generally clear evaluation criteria and use cases for protocol
languages that would enable us to evaluate them on conceptual grounds.
And yet, there is a general belief in the research community that
\review{D:scope1}
existing languages can largely tackle the challenges of building
multiagent systems.
\review{E:scope1}

\subsection*{Contributions, Novelty, and Significance}

We posit that current languages, illustrating the established
paradigms, largely do not support the engineering of decentralized
multiagent systems.
\review{D:scope2}
In support, we contribute a conceptual evaluation of protocol
languages with respect to decentralization.
\review{E:scope2}
More specifiically, our contributions in this paper are the following.

First, we provide evaluation criteria for protocol languages that are
informed by decentralization.  The criteria have to do with how well a
language supports specifying flexible interactions; whether a language
enables expressing the appropriate information constraints; and the
assumptions (demands) a language makes of a MAS's operational
environment.

Second, we undertake a comparative evaluation of selected protocol
languages against the aforementioned criteria.  The selected languages
are modern, diverse, and prominent.  An important feature of our
evaluation is our reliance on ``minimal'' use cases for protocols that
help bring forth the distinctions between the languages.
Specifically, for each criterion, we take realistic use cases that go
to its heart and specify them as best possible in each of the selected
languages.  We then analyze each resulting protocol specification for
validity according to the semantics of the language it is specified
in.  Following this methodology, we show that several of the selected
languages fall short of what is required to support decentralization.

Third, we identify the architectural assumptions underlying the
selected languages and discuss how they map to MAS architecture.

Fourth, we posit principles for engineering MAS and discuss how the
languages fare against them.  \emph{No unitary perspective} states
that a protocol must not specify orderings of events from a unitary
perspective.  \emph{Noninterference} and the \emph{end-to-end
  principle for protocols} both concern layering.  Noninterference
states that a protocol must not interfere with agent reasoning.  The
end-to-end principle states that a protocol can be fully and correctly
implemented only in agents, not in the infrastructure.
\review{B:E2E}Relying on infrastructure for correctness (e.g., via
FIFO message delivery) may be inadequate and unnecessary for
correctness.  The end-to-end principle for protocols derives from the
more general end-to-end principle for system design
\citep{Saltzer+84}.

This paper's \emph{significance} lies in bringing forth the
information models, semantics, and architectural assumptions
underlying protocol languages as they relate to decentralization.
Doing so not only provides a conceptual framework for understanding
protocols and protocol languages but also clarifies requirements for
MAS and yields guidance on research into protocol languages.  Its
\emph{novelty} arises from the absence, currently, of such a
framework.  Notably, this paper focuses on essential representational
criteria for protocols and plays down contingent features such as
current tool support and popularity.

We use the following notational conventions throughout (except in
listings and figures).  We use \rname{small caps} for role names;
\mname{Slant} for protocol and message names; \paraname{sans serif}
for parameter names; and \val{teletype} for parameter values.

\subsection*{Organization}
The rest of this paper is organized as follows.  Section~\ref{sec:MAS}
introduces protocols as an architectural abstraction for MAS.
Section~\ref{sec:selected} introduces the selected approaches we
analyze at depth in this paper.  The paper then motivates each
criterion and studies how the selected languages
fare. Section~\ref{sec:flexibility} evaluates the languages for
concurrency and extensibility, as important aspects of flexibility;
Section~\ref{sec:information} for protocol instances, integrity, and
social meaning, as important aspects related to the information
exchanged in protocol enactments; and Section~\ref{sec:operational}
for assumptions about the operational environment for protocol
enactment.  Section~\ref{sec:mapping} teases out the architectural
models underlying the various languages and compares them to the
canonical MAS architecture, as presented earlier.  It also presents
broad principles for protocol languages and evaluates the selected
languages against them.  Section~\ref{sec:discussion} summarizes the
overall evaluation in the context of alternative evaluations in the
literature. It ends with a discussion of future directions.

\section{Multiagent Systems Architecture}
\label{sec:MAS}

A protocol is an architectural abstraction and therefore any
evaluation of protocol languages must start with a clear understanding
of MAS architecture.  Below, we present a canonical architectural
style \citep{shaw:architecture:1996} for MAS, clearly indicating its
components, assumptions, and constraints.  \review{B:style}The idea is
that any concrete instantiation of the architecture must satisfy
constraints but without making stronger assumptions.

Strictly speaking, agents and roles are distinct categories (an agent
may play several roles and a role may be played by several agents).
For expository convenience, and to focus on other concerns in this
paper, we assume that any role is played by a single agent and
distinct roles are played by distinct agents.  This assumption enables
us to identify an agent with the role it plays.  From here on, we talk
primarily of agents and deemphasize roles.

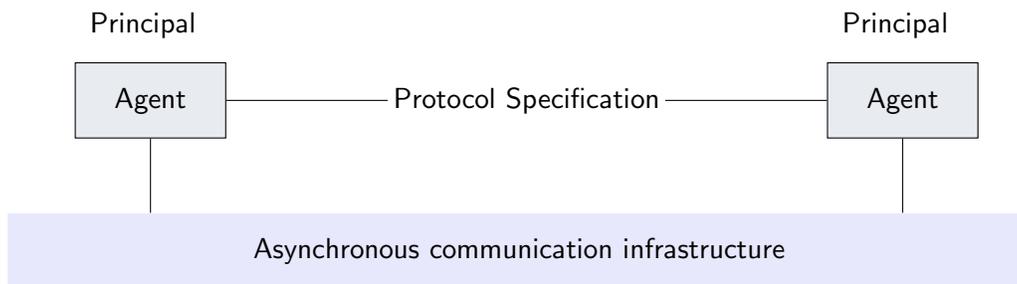
\begin{figure}[ht!]
  \centering
  \begin{tikzpicture}[>=stealth]

\tikzstyle{box}=[draw=none,rounded corners,align=center,font=\sffamily,fill=blue!20!gray!80,rectangle,anchor=center,minimum height=4ex,minimum width=12ex,inner sep=2] %,text width=13ex

\tikzstyle{pbox}=[draw=none,sharp corners,align=center,font=\sffamily,fill=blue!20!gray!40,rectangle,anchor=center,minimum height=6ex,minimum width=20ex,inner sep=2]

\tikzstyle{cbox}=[sharp corners,align=center,font=\sffamily,fill=blue!80!gray!10,rectangle,anchor=center,minimum height=6ex,minimum width=82ex,inner sep=2]

\tikzstyle{ebox}=[draw=none,sharp corners,align=center,font=\sffamily,fill=blue!20!gray!40,rectangle,anchor=center,minimum height=3.5ex,minimum width=12ex,inner sep=2]

\tikzstyle{dmbox}=[draw=none,sharp corners,align=center,font=\sffamily,fill=blue!20!gray!40,rectangle,anchor=center,minimum height=15ex,minimum width=6ex,inner sep=2]

\tikzstyle{lane}=[draw=none]

\tikzstyle{bar}=[draw=none,fill=black,minimum width=8ex]

\tikzstyle{edge_label}=[draw=none, pos=0, fill=white,inner sep=2,font=\sffamily,align=center]
\tikzstyle{west_label}=[edge_label] %,anchor=west
\tikzstyle{east_label}=[edge_label] %,anchor=east

\tikzstyle{edge_label_org}=[draw=none, pos=0, fill=white,inner
sep=2,font=\sffamily] % ,align=center,y

\tikzstyle{every text node part/.style}=[align=center]

\tikzset{
  big arrow/.style={
    decoration={markings,mark=at position 1 with
%% {\arrow[scale=2,#1]{>}}
{\arrow[line width=2.5,#1]{>}}
},
    postaction={decorate},
    shorten >=0.4pt},
  big arrow/.default=blue}

\tikzset{
  big arrow LR/.style={
    decoration={markings,
mark=at position 6pt with {\arrow[line width=2.5,#1]{<}};,
mark=at position 1 with {\arrow[line width=2.5,#1]{>}};},
    postaction={decorate},
    shorten <=1pt,
    shorten >=0.4pt},
  big arrow LR/.default=blue}

\tikzset{darkarrow/.style={big arrow,colorEntityBack,thick}, big arrow/.default=colorEntityBack}

\tikzset{bluearrowLR/.style={big arrow LR,blue, thick}, big
  arrow LR/.default=blue}

\node[pbox,fill=listbackgroundcolorlight] (la) at (-1,0) [draw,minimum width=2cm,minimum height=1cm] {Agent};
\node[pbox,fill=listbackgroundcolorlight] (ra) at (9,0) [draw,minimum width=2cm,minimum height=1cm] {Agent};
\node[box,fill=none] (lp) at ($(la.north)+(-0.1,.5)$) {Principal};
\node[box,fill=none] (rp) at ($(ra.north)+(-0.1,.5)$) {Principal};

\draw  (la)--node [edge_label_org,midway, align=center] (protocol) {Protocol Specification} (ra);

\node[cbox](cic) at  (3.9,-2) {Asynchronous communication infrastructure};
\draw (la) to ($(cic.north)+(-4.9,0)$);
\draw (ra) to ($(cic.north)+(5.1,0)$);
\end{tikzpicture}

\caption{Minimal MAS architecture.  Agents implement the protocol and
  communicate via asynchronous messaging.  The communication
  infrastructure provides no message delivery guarantees other than
  that it is noncreative (delivers only those messages that were sent).}
\label{fig:minimal-MAS}
\end{figure}

As Figure~\ref{fig:minimal-MAS} depicts, each agent represents an
autonomous \emph{principal}, for example, a human or an organization.
An agent internally encodes the private decision making of its
principal, including any private knowledge bases that it relies upon
for decision making.  We elide principals in the later figures.

\review{B:operational}
In general, to achieve interoperation, the interoperating parties must
agree at multiple levels \citep{SOC-05}.  Here, the protocols focus on
the operational level, which concerns the exchange of information
(i.e., reflected in constraints on the ordering and occurrence of
messages).  In particular, we set aside both low-level concerns such
as how the information is encoded and high-level concerns as to the
meaning of the information exchanged---though we insist that protocols
be able to support a flexible representation of meaning.

An agents sends and receives messages via a communication
infrastructure.  An agent's \emph{observations} are its message
emissions and receptions.  For simplicity and in accordance with the
literature, we assume that agents make observations serially
\citep{Hewitt77,Agha86,Fagin+95}.  An agent's \emph{history} is the
set of its observations.  Technically, an agent complies with a
protocol if and only if all of its observations are correct with
respect to the protocol.  Constraints~\ref{req:emission-correctness}
and~\ref{req:reception-correctness} addresses the correctness of
emissions and receptions, respectively.

\begin{constraint}[Emission correctness]\label{req:emission-correctness}
  The correctness of a message emission by an agent may be determined
  from the agent's history.
\end{constraint}

Constraint~\ref{req:emission-correctness} rules out reliance on any
kind of state other than the history for purposes of determining the
correctness of emissions.  In particular, it rules out reliance on (1)
the global state, which may include what the agent has not observed;
(2) the future state of an agent, because any decision should respect
causation; and (3) an agent's internal state, for example, as encoded
in its beliefs \citep{Computer-98}.

\begin{constraint}[Reception correctness]\label{req:reception-correctness}
  The reception of any message that was emitted correctly is correct.
\end{constraint}

\review{B:correctness}

Constraint~\ref{req:reception-correctness} captures the intuition of
respecting the structure of causality.  Specifically, if the reception
of a message would be incorrect, then that message ought never to have
been sent.  Otherwise, the recipient would enter a ``corrupted'' state
from which there is no recourse.  The only alternative would be for
the infrastructure to intervene and prevent a message reception that
would be erroneous, but doing so would customize the infrastructure to
include application-specific details, in contravention of the famous
end-to-end principle \citep{Saltzer+84}, which advocates generality in
the infrastructure.

Constraints~\ref{req:emission-correctness} and
\ref{req:reception-correctness} imply that agents need no more than
an asynchronous communication infrastructure, as captured by
Constraints~\ref{req:nonblocking-emission}
and~\ref{req:anytime-reception}.

\begin{constraint}[Asynchrony: Nonblocking
  emission]\label{req:nonblocking-emission}
  Sends are nonblocking, meaning that when an agent sends a message,
  it does not synchronize with the receiver on the sending action.
\end{constraint}

\begin{constraint}[Asynchrony: Anytime reception]\label{req:anytime-reception}
  An agent receives a message when it is delivered by the
  infrastructure.  That is, message reception is nondeterministic.
\end{constraint}

Of course, an agent being autonomous may choose not to act on a
message it has received but the reception itself occurs due to the
infrastructure.

Asynchrony promotes loose coupling between agents.  Notably,
programming paradigms for building distributed systems such as the
actor model \citep{Hewitt+73,Hewitt77,Agha86} give prominence to
asynchronous messaging for organizing decoupled computations.
Practical communication infrastructures such as the Internet support
asynchronous messaging.  In fact, asynchrony is the only viable option
in the important setting of the Internet of Things (IoT)
\citep{XMPP-15:IoT,MQTT-14,CoAP-14}.

\review{B:assumption}
\begin{assumption}[Infrastructure
  guarantees]\label{req:infrastructure} The infrastructure is noncreative; that is, only sent messages are
  received.  Further, the infrastructure does not deliver corrupt
  messages.
\end{assumption}

Notice that Assumption~\ref{req:infrastructure} does not say that a
sent message be also received.  Indeed, messages may be lost in
transit.  Some applications of MAS may require messages to be
delivered; the analysis in this paper however does not rely upon such
an assumption.  Also notice that no message delivery order was
assumed.  Constraint~\ref{req:reception-correctness} means no such
assumption is required for purposes of correctness.
% Constraint~\ref{req:anytime} does not rule out the
% possibility that an agent might process received messages in an order
% different from the order in which they were received.

% We reserve the term \emph{compliance} (with a protocol) as a property
% of agents; we use the term \emph{correctness} (with respect to a
% protocol) for individual observations.

% A \emph{protocol state} is a snapshot of the histories of all agents
% in the MAS.  A protocol state is purely conceptual; there is no global
% store of protocol state.  In fact, there is no need for such a store
% for purporse of correctness, since correctness is a local
% determination.

\begin{constraint}[Fullness]\label{req:fullness}
  A protocol fully specifies a multiagent system at the operational
  level.
\end{constraint}

Conceptually, as stated above, a protocol specifies the constraints on
the operational level, i.e., on the information exchange.
\review{B:completely}
Constraint~\ref{req:fullness} states that nothing else besides a
protocol is needed to characterize a MAS at the operational level.
That is, this constraint rules out reliance on extra-protocol
mechanisms, such as agreements about when to send or not send certain
messages.  Such extra-protocol mechanisms would amount to hidden
coupling between the agents: agents who were developed to interoperate
in accordance with such mechanisms would not interoperate with agents
who were developed to in accordance solely with the protocol.
Constraint~\ref{req:fullness} means that Figure~\ref{fig:minimal-MAS}
captures a MAS fully from the standpoint of coordination.

% The significant import of Constraint~\ref{req:reception-compliance} is
% that it rules out reliance on message ordering guarantees from the
% communication infrastructure for correctness.

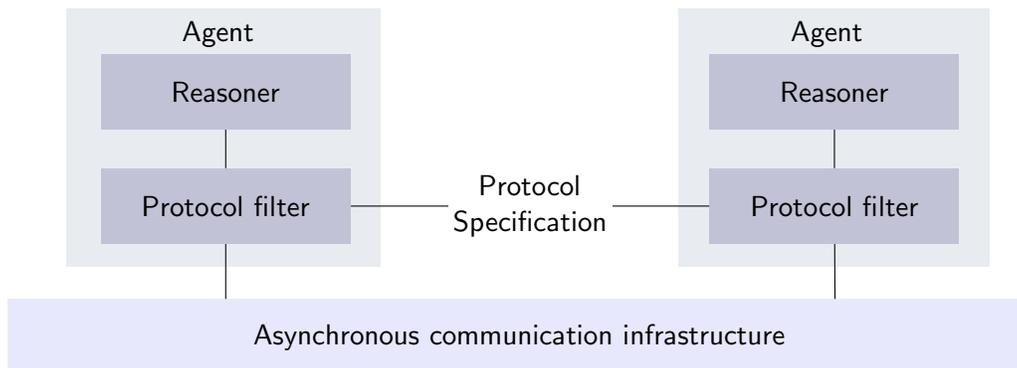
\begin{figure}[ht!]
  \centering
  \begin{tikzpicture}[>=stealth]

\tikzstyle{box}=[draw=none,rounded corners,align=center,font=\sffamily,fill=blue!20!gray!80,rectangle,anchor=center,minimum height=4ex,minimum width=12ex,inner sep=2] %,text width=13ex

\tikzstyle{pbox}=[draw=none,sharp corners,align=center,font=\sffamily,fill=blue!20!gray!40,rectangle,anchor=center,minimum height=6ex,minimum width=20ex,inner sep=2]

\tikzstyle{cbox}=[sharp corners,align=center,font=\sffamily,fill=blue!80!gray!10,rectangle,anchor=center,minimum height=6ex,minimum width=82ex,inner sep=2]

\tikzstyle{ebox}=[draw=none,sharp corners,align=center,font=\sffamily,fill=blue!20!gray!40,rectangle,anchor=center,minimum height=3.5ex,minimum width=12ex,inner sep=2]

\tikzstyle{dmbox}=[draw=none,sharp corners,align=center,font=\sffamily,fill=blue!20!gray!40,rectangle,anchor=center,minimum height=15ex,minimum width=6ex,inner sep=2]

\tikzstyle{lane}=[draw=none]

\tikzstyle{bar}=[draw=none,fill=black,minimum width=8ex]

\tikzstyle{edge_label}=[draw=none, pos=0, fill=white,inner sep=2,font=\sffamily,align=center]
\tikzstyle{west_label}=[edge_label] %,anchor=west
\tikzstyle{east_label}=[edge_label] %,anchor=east

\tikzstyle{edge_label_org}=[draw=none, pos=0, fill=white,inner
sep=2,font=\sffamily] % ,align=center,y

\tikzstyle{every text node part/.style}=[align=center]

\tikzset{
  big arrow/.style={
    decoration={markings,mark=at position 1 with
%% {\arrow[scale=2,#1]{>}}
{\arrow[line width=2.5,#1]{>}}
},
    postaction={decorate},
    shorten >=0.4pt},
  big arrow/.default=blue}

\tikzset{
  big arrow LR/.style={
    decoration={markings,
mark=at position 6pt with {\arrow[line width=2.5,#1]{<}};,
mark=at position 1 with {\arrow[line width=2.5,#1]{>}};},
    postaction={decorate},
    shorten <=1pt,
    shorten >=0.4pt},
  big arrow LR/.default=blue}

\tikzset{darkarrow/.style={big arrow,colorEntityBack,thick}, big arrow/.default=colorEntityBack}

\tikzset{bluearrowLR/.style={big arrow LR,blue, thick}, big
  arrow LR/.default=blue}

\matrix (swiml) [draw=none,matrix of nodes,fill=none,rounded corners,row sep=15,
    column sep=-\pgflinewidth] {
    \node[ebox,fill=none]  {};  \\
      \node[pbox] (dml) {Reasoner};  \\
      \node[pbox] (pcl) {Protocol filter}; \\
  \node[ebox, fill=none] (epl) {};  \\
};

\matrix (swimr) at ($(swiml.east) + (6.3,0.0)$) [draw=none,matrix of nodes,fill=none,rounded corners,row sep=15,
    column sep=-\pgflinewidth] {
      \node[ebox,fill=none] {};\\
   \node[pbox] (dmr) {Reasoner}; \\
   \node[pbox] (pcr) {Protocol filter}; \\
     \node[ebox, fill=none] (epr) {}; \\
};

% \node[dmbox](dmla) at  (-7.3,0.6) {Decision-making};
\draw  (dml) to (pcl);
\draw  (dmr) to (pcr);

 \draw  (pcl)--node [edge_label_org,midway, align=center] (protocol) {Protocol\\Specification} (pcr);

 \node[box,fill=none] (lp) at ($(dml.north)+(-0.1,1)$) {};
 \node[box,fill=none] (rp) at ($(dmr.north)+(-0.1,1)$) {};
\node[box,fill=none](lag) at  ($(lp.south)+(0,-0.4)$) {Agent};
\node[box,fill=none](rag) at  ($(rp.south)+(0,-0.4)$) {Agent};

\node[cbox](cic) at  (3.9,-2.5) {Asynchronous communication infrastructure};
\draw (pcl) to ($(cic.north)+(-3.9,0)$);
\draw (pcr) to ($(cic.north)+(4.2,0)$);

\begin{pgfonlayer}{background}
\draw[draw=none,fill=listbackgroundcolorlight] ($(dml.north west)+(-0.45,0.6)$)  rectangle ($(pcl.south east)+(0.4,-0.3)$);
\draw[draw=none,fill=listbackgroundcolorlight] ($(dmr.north west)+(-0.4,0.6)$)  rectangle ($(pcr.south east)+(0.4,-0.3)$);
\end{pgfonlayer}{background}

\end{tikzpicture}

\caption{MAS architecture with compliance checking.  History is
  maintained by the protocol filter for purposes of compliance checking.}
\label{fig:filter-MAS}
\end{figure}

Figure~\ref{fig:filter-MAS} elaborates on the architecture of
Figure~\ref{fig:minimal-MAS} by refining an agent into two components:
\emph{protocol filter} and \emph{reasoner}.

An agent's protocol filter ensures compliance.  The filter interfaces
with the communication infrastructure to send and receive messages.
And it interfaces with the reasoner to notify the reasoner of
observations of interest and to accept message emission requests.  The
filter materializes the agent's history and uses it to check for
correctness any message that the reasoner requests it to send.  If the
message is correct, the filter sends the message on the infrastructure
(and records the emission as an observation in the history).  If the
message is not correct, the filter discards the message with the
indication of an exception to the reasoner (and does not change the
history).  The filter is a form of generic protocol-based
\emph{control} on the reasoner
\citep{banihashemi:control:2016,banihashemi:control:2018}.

In principle, the filter may have to deal with the incorrect reception
of a message if the message were incorrectly sent.  For simplicity,
let's assume that agents do not send incorrect messages, even when
they are not equipped with a filter.

An agent's reasoner encodes the decision making of its principal.  The
reasoner determines how an agent processes events, both private (e.g.,
an update to an internal knowledge base) and observations recorded by
the filter.  The processing of an event may require the emission of a
message, for which the reasoner relies on the filter.  For example,
referring to \pr~\ref{pr:purchase}, \rname{buyer}'s reasoner, upon
being notified by an internal database that a particular
\paraname{item} was out of stock, may ask the filter to send a
\mname{Request} for that \paraname{item} to \rname{seller}.  If the
\mname{Request} is correct, the filter sends it to \rname{seller}.
\rname{seller}'s reasoner, when notified by its filter of the
reception of the \mname{Request}, may determine by looking up its
internal price list, that an \mname{Offer} for the requested
\paraname{item} should be sent for some \paraname{price}.  And so on.

\begin{figure}[ht!]
  \centering
  \begin{tikzpicture}[>=stealth]

\tikzstyle{box}=[draw=none,rounded corners,align=center,font=\sffamily,fill=blue!20!gray!80,rectangle,anchor=center,minimum height=4ex,minimum width=12ex,inner sep=2] %,text width=13ex

\tikzstyle{pbox}=[draw=none,sharp corners,align=center,font=\sffamily,fill=blue!20!gray!40,rectangle,anchor=center,minimum height=6ex,minimum width=20ex,inner sep=2]

\tikzstyle{cbox}=[sharp corners,align=center,font=\sffamily,fill=blue!80!gray!10,rectangle,anchor=center,minimum height=6ex,minimum width=82ex,inner sep=2]

\tikzstyle{ebox}=[draw=none,sharp corners,align=center,font=\sffamily,fill=blue!20!gray!40,rectangle,anchor=center,minimum height=3.5ex,minimum width=12ex,inner sep=2]

\tikzstyle{dmbox}=[draw=none,sharp corners,align=center,font=\sffamily,fill=blue!20!gray!40,rectangle,anchor=center,minimum height=15ex,minimum width=6ex,inner sep=2]

\tikzstyle{lane}=[draw=none]

\tikzstyle{bar}=[draw=none,fill=black,minimum width=8ex]

\tikzstyle{edge_label}=[draw=none, pos=0, fill=white,inner sep=2,font=\sffamily,align=center]
\tikzstyle{west_label}=[edge_label] %,anchor=west
\tikzstyle{east_label}=[edge_label] %,anchor=east

\tikzstyle{edge_label_org}=[draw=none, pos=0, fill=white,inner
sep=2,font=\sffamily] % ,align=center,y

\tikzstyle{every text node part/.style}=[align=center]

\tikzset{
  big arrow/.style={
    decoration={markings,mark=at position 1 with
%% {\arrow[scale=2,#1]{>}}
{\arrow[line width=2.5,#1]{>}}
},
    postaction={decorate},
    shorten >=0.4pt},
  big arrow/.default=blue}

\tikzset{
  big arrow LR/.style={
    decoration={markings,
mark=at position 6pt with {\arrow[line width=2.5,#1]{<}};,
mark=at position 1 with {\arrow[line width=2.5,#1]{>}};},
    postaction={decorate},
    shorten <=1pt,
    shorten >=0.4pt},
  big arrow LR/.default=blue}

\tikzset{darkarrow/.style={big arrow,colorEntityBack,thick}, big arrow/.default=colorEntityBack}

\tikzset{bluearrowLR/.style={big arrow LR,blue, thick}, big
  arrow LR/.default=blue}

\matrix (swiml) [draw=none,matrix of nodes,fill=none,rounded corners,row sep=15,
    column sep=-\pgflinewidth] {
%
%  \node[box,fill=none] (la) {Principal}; & &
 % \node[box,fill=none] (lr) {Principal}; \\
%
%  \node[pbox] (dml) {Decision making}; & &
 % \node[pbox] (dmr) {Decision making}; \\
%
 %
      \node[ebox,fill=none]  {};  \\
      \node[pbox] (dml) {Reasoner};  \\
      \node[pbox] (mcl) {Meaning computer}; \\
     \node[pbox] (pcl) {Protocol filter}; \\
  \node[ebox, fill=none] (epl) {};  \\
};

\matrix (swimr) at ($(swiml.east) + (6.3,0.0)$) [draw=none,matrix of nodes,fill=none,rounded corners,row sep=15,
    column sep=-\pgflinewidth] {
      \node[ebox,fill=none] {};\\
   \node[pbox] (dmr) {Reasoner}; \\
   \node[pbox] (mcr) {Meaning computer}; \\
      \node[pbox] (pcr) {Protocol filter}; \\
     \node[ebox, fill=none] (epr) {}; \\
};

% \node[dmbox](dmla) at  (-7.3,0.6) {Decision-making};
\draw  (mcl) to (pcl);
\draw  (mcr) to (pcr);
\draw  (dml) to (mcl);
\draw  (dmr) to (mcr);

 \draw  (mcl)--node [edge_label_org,midway, align=center] (meaning) {Meaning\\Specification} (mcr);
 \draw (pcl)--node [edge_label_org,midway,align=center] (protocol) {Protocol\\Specification} (pcr);

%\node[box,fill=none] (lp) at ($(dml.north)+(-0.1,1)$) {Principal};
\node[box,fill=none](lag) at  ($(dml.north)+(0,0.3)$) {Agent};

%\node[box,fill=none] (rp) at ($(dmr.north)+(0.1,1)$) {Principal};
\node[box,fill=none](rag) at  ($(dmr.north)+(0,0.3)$) {Agent};

\node[cbox](cic) at  (3.9,-3.3) {Asynchronous communication infrastructure};
\draw (pcl) to ($(cic.north)+(-3.9,0)$);
\draw (pcr) to ($(cic.north)+(4.2,0)$);

\begin{pgfonlayer}{background}
\draw[draw=none,fill=listbackgroundcolorlight] ($(dml.north west)+(-0.45,0.6)$)  rectangle ($(pcl.south east)+(0.4,-0.3)$);
\draw[draw=none,fill=listbackgroundcolorlight] ($(dmr.north west)+(-0.4,0.6)$)  rectangle ($(pcr.south east)+(0.4,-0.3)$);
\end{pgfonlayer}{background}

\end{tikzpicture}

\caption{Multiagent System Architecture.  Agents interoperate on the basis of protocols and high-level meanings.  Each agent features a protocol filter, a meaning computer, and a reasoner.  A communication infrastructure transports messages between agents.}
\label{fig:meaning-MAS}
\end{figure}
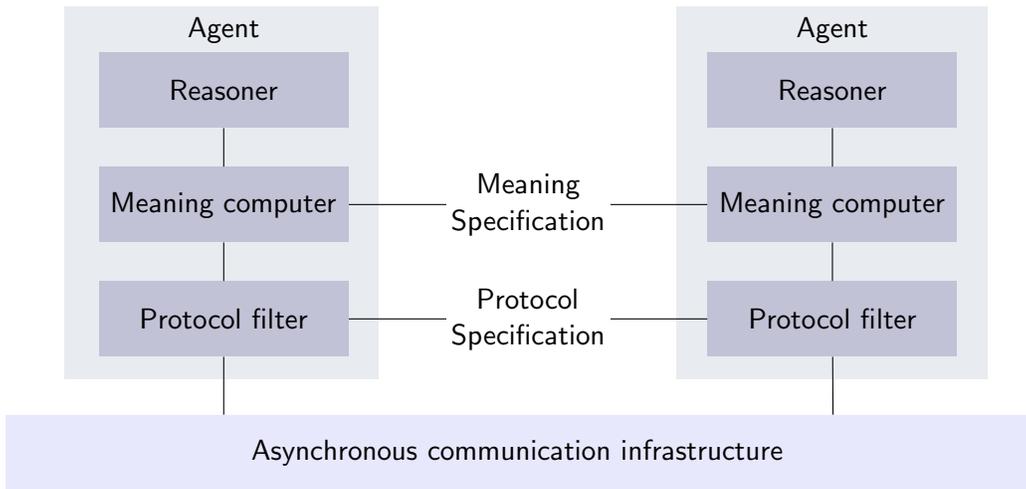

The architecture in Figure~\ref{fig:meaning-MAS} further refines the
architecture in Figure~\ref{fig:filter-MAS} by introducing a
declarative specification of the \emph{social meaning} of an interaction
\citep{Computer-98} and a runtime for the language in which
meaning is specified, namely, the \emph{meaning computer}.  The
meaning specification takes an agent's observations as the base-level
social events and maps combinations of social events to higher-level
social events.

Constraint~\ref{req:social} defines what may be considered a social
event.

\begin{constraint}[Social]\label{req:social}
  A social event is either an observation or is inferred from other
  social events \citep{chopra:iose:2016}.
\end{constraint}

Constraint~\ref{req:social} means that a social event cannot feature
any information that does not show up in an observation (of a message,
as defined earlier).  Internal events that reflect updates to an
agent's internal state (e.g., its beliefs) have no effect on the
computation of social events \citep{Computer-98}.

Social meaning is essential to the application-specific correctness of
MAS.  A MAS for financial loans may model social meaning via
abstractions for \emph{debt}, \emph{collateral}, \emph{default}, and
so on.  For example, from events corresponding to the issuance of a
loan and a payment against the loan, a new debt event could be
inferred that reflects the outstanding debt.  Further, were the
outstanding debt to be zero, it could lead to the inference of a new
\emph{repaid} event.  In like manner, a MAS that supports a community
of toy train enthusiasts could model social meaning via abstractions
for the \emph{provenance}, \emph{ownership}, and \emph{desirability}
of a toy train.

In MAS research, social meaning is often modeled via commitments
\citep{Aamas-Protocols-02,Fornara+Colombetti-02,bentahar:arguments:2004,Winikoff+Liu+Harland-05,Dastani+17:commitments,JAAMAS-18:GoCo,JAIR-19:goals+commitments}
and other norms
\citep{Artikis+Sergot+Pitt-TOCL-09,padget:deontic-sensors:2018,alechina:norms:2018}.
In the rest of the paper, for reasons of concreteness and familiarity,
we use commitments as an exemplar way of modeling social meaning.
\pr~\ref{pr:commitment} illustrates the use of commitments to
capture meaning.

\begin{example}[Deliver-Payment Commitment]\label{pr:commitment}
  In the context of purchase in \pr~\ref{pr:purchase}, the meaning of
  an \mname{Accept} from \rname{seller} to \rname{buyer} for some
  \paraname{item} for some \paraname{price} is that it creates a
  commitment from \rname{buyer} to \rname{seller} that if
  \rname{seller} \mname{Deliver}s the \paraname{item} by some
  (specified) deadline, then \rname{buyer} will make a \mname{Payment}
  of the \paraname{price} by some (specified) deadline.
\end{example}

\section{Overview of Selected Approaches}
\label{sec:selected}

For this evaluation, we select protocol specification languages that
are recent, have a formal semantics, and represent diverse doctrines.
\review{B:AUML-a} AUML notably is not in our selection: neither is it
recent nor does it have a satisfactory formal semantics.  AUML is
closely related to UML Sequence Diagrams, which too lacks a formal
semantics.  Some of the selected languages though adopt important
intuitions behind AUML, including the idea of specifying an
interaction as a control flow of messages and using a graphical
notation.  One might argue that some of the languages we discuss
(Scribble and Trace-F) seek to clarify intuitions that undergird AUML
and to provide a formal semantics.

Below, we introduce the main ideas of the selected languages by
specifying \pr~\ref{pr:purchase}.

\subsection{Multiparty Session Types: Scribble}
Scribble \citep{Yoshida+13:Scribble} is a practical instantiation of
multiparty session types \citep{Honda+16:JACM}.  In Scribble, a
protocol is an ordering of constituent protocols (bottoming out at
individual message specifications) using constructs such as sequence,
choice, and recursion.  Scribble assumes that communication between
pairs of participants is asynchronous but ordered over FIFO channels.

Listing~\ref{Scribble:Purchase} gives an encoding of
\pr~\ref{pr:purchase} as a Scribble protocol.  In the listing, a
semicolon (\fbf{;}) indicates sequencing.

\begin{lstlisting}[caption={\mname{Purchase} (\pr~\ref{pr:purchase}) in Scribble.},label={Scribble:Purchase}]
global protocol Purchase(role Buyer, role Seller) {
 Request() from Buyer to Seller;
 Offer() from Seller to Buyer;

 choice at Buyer {
   Accept() from Buyer to Seller;
   Deliver() from Seller to Buyer;
   Payment() from Buyer to Seller;
  } or {
   Reject() from Buyer to Seller;
  }
}
\end{lstlisting}

Given a protocol, Scribble yields projections, called \emph{local
  protocols}, for each agent.  (We retain the term ``projection'' to
avoid conflict with ``protocol.'')  The idea is that the protocol
represents computations from a unitary perspective whereas an agent's
projection represents computations from its own local perspective.
Scribble's tools \citep{scribble-tools} may be used to generate these
projections.  We have used the tooling to verify all Scribble
specifications presented in this paper.

Listing~\ref{Scribble:Purchase-local} gives the projections for each of
the agents in the \pname{Purchase} protocol in
Listing~\ref{Scribble:Purchase}.  \rname{buyer}'s projection says:
send \mname{Request} to \rname{seller}, then receive \mname{Offer}
(from \rname{seller}), then send either \mname{Accept} or
\mname{Reject}.  If \mname{Accept} is sent, then receive
\mname{Deliver} and then send \mname{Payment}.  \rname{seller}'s
projection is read in an analogous manner.

Notice that in the protocol the choice between \mname{Accept} and
\mname{Reject} is indicated as \rname{buyer}'s.  Therefore, in the
projections, the choice is interpreted as an internal choice for
\rname{buyer} and as an external choice for \rname{seller}.  The agent
with an internal choice chooses from the available alternatives
autonomously.  The agent with an external choice does not choose but
follows along.  The internal choice determines the external choice.
Thus, if \rname{buyer} chooses to send \mname{Accept} (alternatively,
\mname{Reject}), its reception resolves the \rname{seller}'s choice to
receive \mname{Accept} (alternatively, \mname{Reject}).

\begin{lstlisting}[caption={Scribble projections of \mname{Purchase} (Listing~\ref{Scribble:Purchase}) for \rname{buyer} and \rname{seller}.},label={Scribble:Purchase-local}]
local protocol Purchase_Buyer(role Buyer, role Seller) {
 Request() to Seller;
 Offer() from Seller;

 choice at Buyer {  //internal choice
   Accept() to Seller;
   Deliver() from Seller;
   Payment() to Seller
 } or {
   Reject() to Seller;
 }
}

local protocol Purchase_Seller(role Buyer, role Seller) {
 Request() from Buyer;
 Offer() to Buyer;

 choice at Buyer {    //external choice
   Accept() from Buyer;
   Deliver() to Buyer;
   Payment() from Buyer;
 } or {
   Reject() from Buyer;
 }
}
\end{lstlisting}

The notion of realizability ties together a protocol and its
projections.  A protocol is \emph{realizable} if and only if the
agents acting locally based on their projections jointly realize
exactly the computations of the protocol (as we shall see, this is not
always the case).  The \mname{Purchase} protocol in
Listing~\ref{Scribble:Purchase} is realizable.

\subsection{Trace-C}

\citet{Castagna+12:sessions} describe a language for specifying
protocols that is based upon trace expressions, which we refer to as
\emph{Trace-C}.  A trace is a sequence of communication events.  In
Trace-C, each expression maps to a set of traces.
\review{B:numbering}The expression $x \ar{m} y$ is atomic; it denotes
the communication of message $m$ from $x$ to $y$; and it maps to the
following set of traces containing just one trace: $\{m\}$.  The
\fbf{;} operator denotes sequential composition; the expression $e;f$
is the concatenation of the traces of $e$ with the traces of $f$.  The
$\mathbf{\lor}$ operator denotes choice; the expression $e\lor f$ is
the union of traces of $e$ and the traces of $f$.  The
$\mathbf{\land}$ operator denotes the shuffle of its operands; the
expression $e\land f$ is the set of those traces that represent an
interleaving of a trace of $e$ with a trace of $f$.  Like Scribble,
Trace-C assumes FIFO-based asynchronous communication.

Listing~\ref{Trace-C:Purchase} shows how \pr~\ref{pr:purchase} may be
rendered in Trace-C.  Although the Trace-C specification appears more
algebraic than Scribble, we can see that they are structurally similar
once we realize that the \fsf{choice} operator in Scribble corresponds
to the $\lor$ operator in Trace-C.

\begin{lstlisting}[label={Trace-C:Purchase},caption={\mname{Purchase} protocol in Trace-C (and Trace-F).}]
 Buyer $\ar{Request}$ Seller ; Seller $\ar{Offer}$ Buyer ;
    ((Buyer $\ar{Accept}$ Seller ; Seller $\ar{Deliver}$ Buyer ; Buyer $\ar{Payment}$ Seller) $\lor$ Buyer $\ar{Reject}$ Seller)
\end{lstlisting}

Like in Scribble, a Trace-C protocol yields projections for each
agent.  Listing~\ref{Trace-C:Purchase-local} gives the projections for
\mname{Purchase} in Listing~\ref{Trace-C:Purchase}.  In the
projections, $\mathbf{\oplus}$, \fbf{+}, and \fbf{;} denote internal
choice, external choice, and sequence, respectively;
\rname{agent}!\mname{Message} and \rname{agent}?\mname{Message},
respectively, denote the emission of \mname{Message} to \rname{agent}
and the reception of \mname{Message} from \rname{agent}.  The
projections are structurally similar to those of \pname{Purchase} in
Scribble, even though the syntax is different.  Notice that
\rname{buyer}'s choice is internal and \rname{seller}'s external,
meaning that although \rname{seller} could receive either
\mname{Accept} or \mname{Reject}, the choice of what it receives
depends on what \rname{buyer} sends.  The protocol is realizable.

\begin{lstlisting}[label={Trace-C:Purchase-local},caption={Trace-C projections of \pname{Purchase} in Listing~\ref{Trace-C:Purchase}.}]
//Buyer's projection
Buyer: Seller!Request ; Seller?Offer ;
  ((Seller!Accept ; Seller?Deliver ; Seller!Payment) $\oplus$ Seller!Reject)

//Seller's projection
Seller: Buyer?Request ; Buyer!Offer ;
  ((Buyer?Accept ; Buyer!Deliver ; Buyer?Payment) $+$ Buyer?Reject)
\end{lstlisting}

\subsection{Trace-F}

\citet{ferrando:enactability:2019} describe a trace expressions-based
language for specifying protocols, which we refer to as
\emph{Trace-F}.  It builds upon earlier work on monitoring
decentralized MAS \citep{ferrando2017decamon}.  Trace-F, like Trace-C,
features operators for sequence, choice, and shuffle.
\review{B:choice}
In Trace-F, shuffle is represented $\mathbf{|}$;
however, for uniformity with Trace-C, we use the Trace-C
representation for shuffle, that is, $\mathbf{\land}$.
\review{C:choice}
With this
simplification, the protocol in Listing~\ref{Trace-C:Purchase} serves
as a specification of \pr~\ref{pr:purchase} in Trace-F.

The projections generated by Trace-F for
Listing~\ref{Trace-C:Purchase} though are different from the
projections produced by Trace-C as shown in
Listing~\ref{Trace-C:Purchase-local}.  Specifically, in Trace-F, the
choice in the protocol does \emph{not} reduce to internal and external
choice in the projections for \rname{buyer} and
\rname{seller}.  Instead, the choice is preserved in the projection and
the distinction between internal and external choice is captured
semantically in a decision structure.  In general, Trace-F preserves
all binary operators used in a protocol in the projections.

\begin{lstlisting}[label={Trace-F:Purchase-local},caption={Trace-F projections of \pname{Purchase} in Listing~\ref{Trace-C:Purchase}.}]
//Buyer's projection
Buyer: Seller!Request ; Seller?Offer ;
 ((Seller!Accept ; Seller?Deliver ; Seller!Payment) $\lor$ Seller!Reject)

//Seller's projection
Seller: Buyer?Request ; Buyer!Offer ;
 ((Buyer?Accept ; Buyer!Deliver ; Buyer?Payment) $\lor$ Buyer?Reject)
\end{lstlisting}

\citet{ferrando:enactability:2019} introduce two dimensions of
variation in reasoning about the realizability (which they term
``enactability'') of a protocol.  One dimension concerns the
communication infrastructure---whether it is asynchronous or
synchronous and if it is asynchronous what kind of ordered delivery
guarantees it offers.  Out of the other approaches evaluated in this
paper that support asynchrony, none requires stronger ordering
guarantees than FIFO delivery.  Hence, the interesting cases for
Trace-F, for our purposes, are asynchrony without any kind of ordered
delivery, which we refer to as \emph{unordered asynchrony}, and
asynchrony with FIFO delivery, which we refer to as \emph{FIFO
  asynchrony}.

The other dimension that \citet{ferrando:enactability:2019}
introduce \review{B:DS}(drawing upon \citep{desai:enactability:2008}) concerns how
the sequence operator is interpreted in terms of the observations of
events.  Take the protocol in Listing~\ref{Trace-F:semantics}.

\begin{lstlisting}[label={Trace-F:semantics},caption={A Trace-F protocol.}]
  W $\ar{p}$ X ; W $\ar{q}$ Y
\end{lstlisting}

Under the \emph{send before send} (SS) interpretation, \rname{w} must
send \mname{p} before \rname{w} sends \mname{q}.  Under the \emph{send
  before receive} (SR) interpretation, \rname{w} must send \mname{p}
before \rname{y} receives \mname{q}.  Under the \emph{receive before
  send} (RS) interpretation, \rname{x} must receive \mname{p} before
\rname{w} sends \mname{q}.  And, under the \emph{receive before
  receive} (RR) interpretation, \rname{x} must receive \mname{p}
before \rname{y} receives \mname{q}.

Whether a protocol is realizable depends on the communication
infrastructure and the interpretation of the sequence operator.  For
concreteness, let's consider the protocol in
Listing~\ref{Trace-F:semantics} under unordered asynchrony.  The
protocol is realizable with either SS (\rname{w} being the sender of
both \mname{p} and \mname{q} can ensure that \mname{p} is sent before
\mname{q}) or SR (from the fact that the protocol is realizable under
SS and the emission of a message must be prior to its reception).
However, the protocol is realizable neither under RS (\rname{w} has no
way of knowing when \rname{x} has received \mname{p}, so it cannot
ensure that \mname{q} will be sent after the reception of \mname{p})
nor under RR (since the receivers are different, there is no way to
ensure that \mname{q} will be received after the reception of
\mname{p}).  Changing the interpretation to FIFO asynchrony makes no
difference (because the receivers of \mname{p} and \mname{q} are
different).

To see how the choice of communication infrastructure makes a
difference, consider the protocol in
Listing~\ref{Trace-F:semantics-FIFO}.  Notice that both \mname{p} and
\mname{q} are messages from \rname{w} to \rname{x}.  Under unordered
asynchrony and with the RR interpretation, the protocol is
unrealizable (there being no way to guarantee that \mname{p} will be
received before \mname{q}).  However, under FIFO asynchrony and the RR
interpretation, the protocol is realizable (\mname{p} is sent before
\mname{q}, so \mname{p} is also received before \mname{q}).

\begin{lstlisting}[label={Trace-F:semantics-FIFO},caption={A Trace-F protocol.}]
  W $\ar{p}$ X ; W $\ar{q}$ X
\end{lstlisting}

\subsection{HAPN}

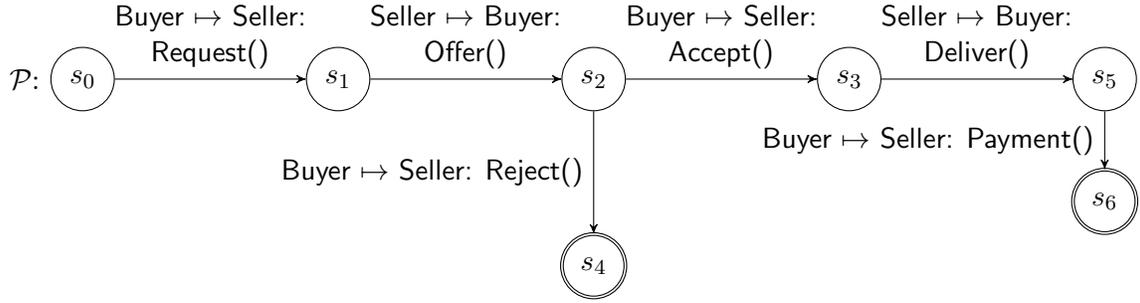
\begin{figure}[htb!]
\centering
\begin{tikzpicture}[>=stealth']

\graph [nodes={circle,inner sep=0.4em,draw}, math nodes,
        edges={align=center,font=\sffamily},
        branch down sep=2em, grow right sep=6.6em,]
{
  s_0/"s_0"-> ["Buyer $\mo$ Seller:\\Request()"] {
    s_1-> ["Seller $\mo$ Buyer:\\Offer()"] {
      s_2-> ["Buyer $\mo$ Seller:\\ Accept()"] s_3-> ["Seller $\mo$ Buyer:\\ Deliver()"] {
        s_5,
        s_5-> ["Buyer $\mo$ Seller: Payment()",swap] s_6[double]
      };
        s_2-> ["Buyer $\mo$ Seller: Reject()",swap] s_4[double,y=2.0em];
    }
  }
};

\node[anchor=east] at (s_0.west) {$\mathcal{P}\mathbf{:}$};

\end{tikzpicture}
\caption{\mname{Purchase} in HAPN, starting from $s_0$.}
\label{hapn:Purchase}
\end{figure}

HAPN \citep{Winikoff+18:HAPN} is a graphical protocol language that
supports nested state machines in a manner similar to
\review{B:statecharts-a} statecharts \citep{Harel-87}.  As
Figure~\ref{hapn:Purchase} shows, nodes represent states or reference
other protocols to compose those protocols.  Edges can have complex
annotations, supporting the specification of message transmissions,
guard expressions, and changes to the state.  HAPN specifies the
enactments of a protocol in terms of state machines.  It assumes
synchronous communication \citep[p.~61]{Winikoff+18:HAPN} and does not
give a method for projecting a protocol to local perspectives (though
\citeauthor{Winikoff+18:HAPN} acknowledge the need to develop such
methods).

HAPN provides methods to flatten a hierarchical protocol into simple protocols and finite state machines for verification.

\subsection{BSPL}
\label{sec:bspl}

BSPL \citep{AAMAS-BSPL-11,AAMAS-BSPL-12}, the Blindingly Simple
Protocol Language, and Splee \citep{AAMAS-17:Splee}, which extends
BSPL, are exemplars of information-based languages.  Instead of
specifying the control flow between messages, BSPL specifies
information causality and integrity constraints.

Listing~\ref{BSPL:Purchase} shows the \pname{Purchase} protocol in
BSPL.  It specifies a set of roles, a set of parameters, and a set of
messages.  In \pname{Purchase}, the roles are \rname{buyer} and
\rname{seller}; the parameters are \paraname{ID}, \paraname{item},
\paraname{price}, \paraname{decision}, and \paraname{OK}; and message
schemas are \mname{Request}, \mname{Offer}, and so on.
\mname{Request} is a message from \rname{buyer} to \rname{seller} and
has parameters \paraname{ID} and \paraname{item}.  BSPL is
declarative; the order in which the messages appear in a protocol is
irrelevant to how the protocol may be enacted.

\begin{lstlisting}[caption={\pname{Purchase} in BSPL.},label={BSPL:Purchase}]
Purchase {
 $\role$ Buyer, Seller
 $\param$ out ID key, out item, out price, out decision, out OK

 Buyer $\mo$ Seller: Request[out ID, out item]
 Seller $\mo$ Buyer: Offer[in ID, in item, out price]
 Buyer $\mo$ Seller: Accept[in ID, in item, in price, out decision, out address]
 Buyer $\mo$ Seller: Reject[in ID, in item, in price, out decision, out OK]
 Seller $\mo$ Buyer: Deliver[in ID, in item, in address, out dropOff]
 Buyer $\mo$ Seller: Payment[in ID, in price, in dropOff, out OK]
}
\end{lstlisting}

A BSPL protocol may be viewed as an information object as described by
the protocol parameters, at least one of which is annotated
$\msf{key}$.  The $\msf{key}$ parameters enable identifying
\emph{instances} of the protocol: distinct bindings for the key
parameters identify distinct \emph{instances} of the protocol.
\pname{Purchase} specifies \paraname{ID} as its key parameter.  A key
parameter of the protocol is also a key parameter of the messages in
which it appears and enables identifying distinct instances of
messages.  Parameter \paraname{ID} is key for all messages in
\pname{Purchase}.  Protocol instances are related to protocol
enactment: A protocol instance is a \emph{view} over correlated (by
bindings of common keys) message instances.  For example, an emission
of \mname{Request} with \paraname{ID} \val{1} and \paraname{item}
\val{fig} yields a protocol instance with \paraname{ID} \val{1} and
\paraname{item} \val{fig}.

A protocol instance must satisfy \emph{integrity}, which is the idea
that no two message instances that are correlated with the protocol
instance may conflict on (that is, have different bindings for) any
parameter---this is the meaning of a key.  Thus for example, a
\mname{Request} with \paraname{ID} \val{1} and \paraname{item}
\val{fig} and an \mname{Offer} with \paraname{ID} \val{1} and
\paraname{item} \val{jam} would violate integrity: \paraname{ID}
\val{1} may either be associated with \paraname{item} \val{fig} or
\paraname{item} \val{jam}, but not both.

For any instance, \emph{causality} constraints specify information
flow and are expressed via $\inn$ and $\out$ adornments on parameters
(we omit discussion of the $\nil$ adornment since it does not feature
in any BSPL specification in the present paper).  Ordering between
messages falls out of these constraints.  To see how, consider
\mname{Request}.  In \mname{Request}, both \paraname{ID} and
\paraname{item} are adorned $\out$, meaning that in sending a
\mname{Request}, \rname{buyer} \emph{produces} bindings for
\paraname{ID} and \paraname{item}.  When \rname{seller} receives the
\mname{Request}, it comes to know those bindings unless it knew them
already.  In \mname{Offer}, both \paraname{ID} and \paraname{item} are
adorned $\inn$, meaning that \rname{seller} needs to \emph{know} these
parameters before \rname{seller} can send \mname{Offer}.  This means
that if \rname{seller} has seen \mname{Request} before, it can send
\mname{Offer} by producing a binding for \paraname{price}.  When
\rname{buyer} receives \mname{Offer}, it may send either
\mname{Accept} or \mname{Reject} since it knows the bindings of
\paraname{ID}, \paraname{item}, and \paraname{price} and it may
produce a binding for \paraname{address} (which features in
\mname{Accept} but not \mname{Reject}), \paraname{decision} and
\paraname{OK} (which features in \pname{Reject} but not
\mname{Accept}).  It cannot send both \mname{Accept} and
\mname{Reject} though because both messages produce a binding for
\paraname{decision}, and integrity requires a parameter to have at
most one binding.  When \rname{seller} receives \mname{Accept}, it
knows \paraname{address} and therefore may send \mname{Deliver} by
producing a binding for \paraname{dropOff}.  Upon reception of
\mname{Deliver}, \rname{buyer} knows \paraname{dropOff}, and
therefore, it can send \mname{Payment} by producing a binding for
\paraname{OK}.

A tuple of bindings for a protocol's parameters corresponds to a
\emph{complete} protocol instance.  That is the motivation for
\mname{Purchase} being designed such that \mname{Reject} features
\paraname{OK} but \mname{Accept} does not.  On the \mname{Accept}
branch, the protocol completes with \mname{Payment}.

\review{C:decen}
Unlike the languages introduced earlier, BSPL does
not give the computations of a protocol from a unitary perspective.
Instead, it takes an inherently decentralized perspective.  Any
computation of a protocol is a vector comprised of a history for each
agent.  However, the vector is conceptual (not materialized anywhere).
To be able to correctly enact a protocol, an agent needs no more than
its history.  Therefore projections are trivial in BSPL.

BSPL works with asynchronous communication without any ordering
guarantees.

\section{Flexibility}
\label{sec:flexibility}

Does a language support specifying flexible protocols?  Being able to
interact flexibly is supportive of autonomy
\citep{Aamas-Protocols-02}.  \review{B:mot1} However, flexibility is
in tension with decentralization: Independently-constructed agents
deciding locally must still be able to interoperate.  Below, we
discuss \emph{concurrency} and \emph{extensiblity}, two aspects of
flexibility.

Below, we denote enactments via sequence diagrams, as in
Figure~\ref{fig:concurrency}, where each agent's lifeline captures its
history.  \review{B:enactment}

\subsection{Concurrency}
\label{sec:concurrency}

\review{B:phrasing} Does a language enable protocols in which agents
may emit and receive messages concurrently?  Consider~\ref{pr:flexible-purchase}.

\begin{example}[Flexible purchase]\label{pr:flexible-purchase}
  \rname{buyer} sends \mname{Request} to \rname{seller} to ship some
  item.  After sending \mname{Request}, \rname{buyer} may send
  \mname{Payment}.  After receiving \mname{Request}, \mname{Seller}
  may send \mname{Shipment}.  That is, \mname{Payment} and
  \mname{Shipment} are not mutually ordered.
\end{example}

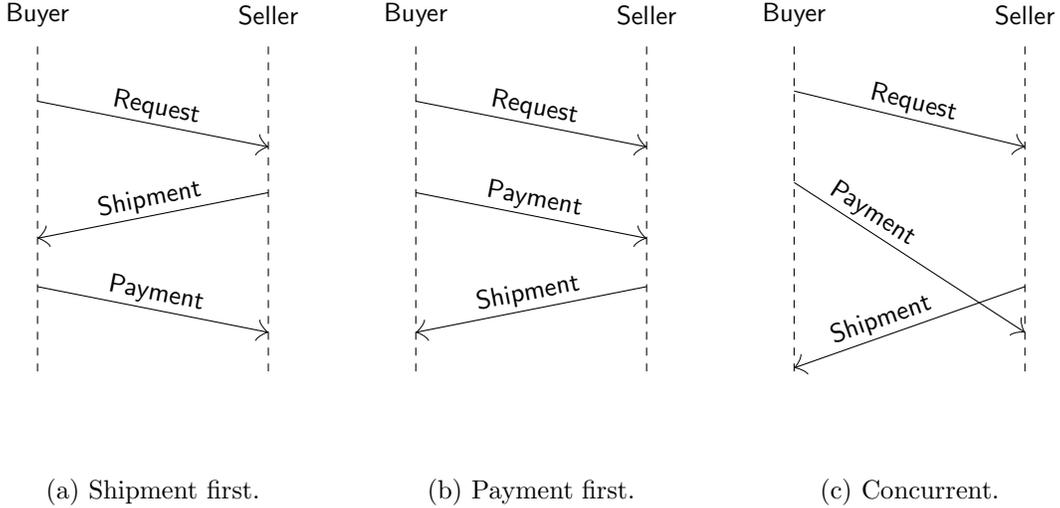
\begin{figure}[htb]
  \tikzstyle{role}=[thin,draw,align=center,font=\small\sffamily,rectangle,anchor=center,minimum height=5ex,minimum width=6ex,inner sep=1]

  \tikzstyle{every text node part/.style}=[align=center]

  \tikzset{every node text/.style={node
      contents=\transform{#1}}}
\newcommand{\transform}[1]{\ensuremath{\mathsf{#1}}}

  \tikzstyle{m_label_base}=[draw=none,midway,fill=none,sloped,align=center,font=\small\sffamily]
  \tikzstyle{m_label_up}=[m_label_base,above=-2pt]
  \tikzstyle{m_label_down}=[m_label_base,below=-2pt]

 \tikzstyle{a_label}=[draw=none,sloped,fill=white,align=center,font=\small\scshape]

% \tikzstyle{message}=[->, >=stealth']
  \tikzstyle{emptybox}=[draw=none,minimum height=3ex]

\centering
\begin{subfigure}[t]{0.33\linewidth}
  \centering
  \begin{tikzpicture}
    \matrix () [row sep=10,column sep=80] {
  \node[emptybox] (a) {};
  &  \node[emptybox] (b) {};\\
  \node (a-zero) {};
  & \node (b-zero) {}; \\
  \node (a-one) {};
  & \node (b-one) {}; \\
  \node (a-two) {};
  & \node (b-two) {}; \\
  \node (a-three) {};
  & \node (b-three) {}; \\[1]
  \node (a-four) {};
  & \node (b-four) {}; \\
  \node (a-five) {};
  & \node (b-five) {}; \\
  \node (a-six) {};
  & \node (b-six) {}; \\
  \node[emptybox] (ae) {};
  &  \node[emptybox] (be) {};\\
};
\node [role,draw=none,anchor=south] at (a) {Buyer};
\node [role,draw=none,anchor=south] at (b) {Seller};

\draw [dashed] (a.center)--(a-six.center);
\draw [dashed] (b.center)--(b-six.center);

\draw [message] (a-zero.center)--node [m_label_up] {Request} (b-one.center);

\draw [message] (b-two.center)--node [m_label_up] {Shipment} (a-three.center);

\draw [message] (a-four.center)--node [m_label_up] {Payment} (b-five.center);

\end{tikzpicture}
\caption{Shipment first.}
\label{fig:shipment-first}
\end{subfigure}% <== Mandatory %
%\hfill
\begin{subfigure}[t]{0.33\linewidth}
  \centering
\begin{tikzpicture}
    \matrix () [row sep=10,column sep=80] {
  \node[emptybox] (a) {};
  &  \node[emptybox] (b) {};\\
  \node (a-zero) {};
  & \node (b-zero) {}; \\
  \node (a-one) {};
  & \node (b-one) {}; \\
  \node (a-two) {};
  & \node (b-two) {}; \\
  \node (a-three) {};
  & \node (b-three) {}; \\[1]
  \node (a-four) {};
  & \node (b-four) {}; \\
  \node (a-five) {};
  & \node (b-five) {}; \\
  \node (a-six) {};
  & \node (b-six) {}; \\
  \node[emptybox] (ae) {};
  &  \node[emptybox] (be) {};\\
};
\node [role,draw=none,anchor=south] at (a) {Buyer};
\node [role,draw=none,anchor=south] at (b) {Seller};

\draw [dashed] (a.center)--(a-six.center);
\draw [dashed] (b.center)--(b-six.center);

\draw [message] (a-zero.center)--node [m_label_up] {Request} (b-one.center);

\draw [message] (a-two.center)--node [m_label_up] {Payment} (b-three.center);

\draw [message] (b-four.center)--node [m_label_up] {Shipment} (a-five.center);
\end{tikzpicture}
\caption{Payment first.}
\label{fig:payment-first}
\end{subfigure}% <== Mandatory %
%\hfill
\begin{subfigure}[t]{0.33\linewidth}
  \centering
\begin{tikzpicture}
    \matrix () [row sep=10,column sep=80] {
  \node[emptybox] (a) {};
  &  \node[emptybox] (b) {};\\
  \node (a-zero) {};
  & \node (b-zero) {}; \\
  \node (a-one) {};
  & \node (b-one) {}; \\
  \node (a-two) {};
  & \node (b-two) {}; \\
  \node (a-three) {};
  & \node (b-three) {}; \\[1]
  \node (a-four) {};
  & \node (b-four) {}; \\
  \node (a-five) {};
  & \node (b-five) {}; \\
  \node (a-six) {};
  & \node (b-six) {}; \\
  \node[emptybox] (ae) {};
  &  \node[emptybox] (be) {};\\
};
\node [role,draw=none,anchor=south] at (a) {Buyer};
\node [role,draw=none,anchor=south] at (b) {Seller};

\draw [dashed] (a.center)--(a-six.center);
\draw [dashed] (b.center)--(b-six.center);

\draw [message] (a-zero.north)--node [m_label_up] {Request} (b-one.center);

\draw [message] (a-two.north)--node [m_label_up,pos=0.3] {Payment} (b-five.center);

\draw [message] (b-four.center)--node [m_label_up,pos=0.6] {Shipment} (a-six.north);
\end{tikzpicture}
\caption{Concurrent.}
\label{fig:crossing}
\end{subfigure}
\hfill
\caption{Three possible enactments of \pname{Purchase}.}
\label{fig:concurrency}
\end{figure}

Figure~\ref{fig:concurrency} shows some possible enactments
for \pr~\ref{pr:flexible-purchase}.

Listing~\ref{Trace-C:concurrency} serves as a protocol specification
in both Trace-C and Trace-F that prima facie captures
\pr~\ref{pr:flexible-purchase} by not mutually ordering
\mname{Payment} and \mname{Shipment}.

\begin{lstlisting}[label={Trace-C:concurrency},caption={Flexible purchase (\pr~\ref{pr:flexible-purchase}) in Trace-C and Trace-F.}]
 //Flexible purchase
Buyer $\ar{Request}$ Seller; (Buyer $\ar{Payment}$ Seller $\land$ Seller $\ar{Shipment}$ Buyer)
\end{lstlisting}

To understand what enactments are supported by the protocol in
Listing~\ref{Trace-C:concurrency}, following Trace-C
\citep[p.~14]{Castagna+12:sessions}, we eliminate $\land$ from the
protocol to obtain the equivalent protocol in
Listing~\ref{Trace-C:concurrency-transform}.  Trace-C determines the
protocol in Listing~\ref{Trace-C:concurrency-transform} as
unrealizable.

\begin{lstlisting}[label={Trace-C:concurrency-transform},caption={A
    Trace-C protocol equivalent to the protocol in
    Listing~\ref{Trace-C:concurrency}.}]
  Buyer $\ar{Request}$ Seller;
  ((Buyer $\ar{Payment}$ Seller ; Seller $\ar{Shipment}$ Buyer) $\lor$ (Seller $\ar{Shipment}$ Buyer; Buyer $\ar{Payment}$ Seller))
\end{lstlisting}

Let's see why.  Listing~\ref{Trace-C:hypo-local-concurrency} gives the
projections that Trace-C yields for
Listing~\ref{Trace-C:concurrency-transform}.  The choice (denoted by
$\lor$) in the protocol must be interpreted as external choice
(denoted $+$) for one agent and internal choice (denoted $\oplus$) for
the other.  An agent with an internal choice can choose autonomously.
An agent with an external choice cannot; its choice is determined by
the internal choice of another agent.  In
Listing~\ref{Trace-C:hypo-local-concurrency}, \rname{buyer} has the
internal choice and \rname{seller} the external choice (it wouldn't
matter to our analysis if it were the other way around since the
situation is symmetric).

\begin{lstlisting}[label={Trace-C:hypo-local-concurrency},caption={Trace-C projections of the protocol in Listing~\ref{Trace-C:concurrency-transform}.}]
Buyer: Seller!Request.
        ((Seller?Shipment.Seller!Payment) $\oplus$ (Seller!Payment.Seller?Shipment))

Seller: Buyer?Request.
        ((Buyer!Shipment.Buyer?Payment) + (Buyer?Payment.Buyer!Shipment))
\end{lstlisting}

Given the projections in Listing~\ref{Trace-C:hypo-local-concurrency},
if \rname{buyer} chooses to send \mname{Payment}, when \mname{Payment}
reaches \rname{seller}, it effectively determines the choice to
receive \mname{Payment} by \rname{seller}.  Such an enactment realizes
the protocol trace where \mname{Payment} happens before
\mname{Shipment}, so no problem here.  However, if \rname{buyer}
chooses to receive \mname{Shipment}, \rname{seller} must send it.  The
\rname{seller} could send \mname{Shipment} if it knew of
\rname{buyer}'s choice or it could act autonomously.  However neither
is a possibility.  Constraint~\ref{req:fullness} rules out covert
communication and synchronization and therefore rules out the
possibility of the \rname{seller} learning of \rname{buyer}'s choice.
As \rname{seller}'s choice is internal, it cannot send
\mname{Shipment} autonomously.  This means the system is deadlocked,
which leads Trace-C to conclude that the protocol is unrealizable.
The situation where agents must make mutually compatible choices to
ensure correctness is known as nonlocal choice
\citep{ladkin:mfg:1995}.

Listing~\ref{Trace-F:local-concurrency} shows the projections in
Trace-F for the protocol in Listing~\ref{Trace-C:concurrency}.  Under
both unordered and FIFO asynchrony, the protocol is determined
unrealizable by Trace-F, no matter what interpretation is chosen for
the sequence operator.  The reason behind the rejection is the same
reason the Trace-C protocol above is rejected: a nonlocal choice that
cannot always be made in a mutually compatible manner by the agents.

\begin{lstlisting}[label={Trace-F:local-concurrency},caption={Projections in Trace-F that illustrate the difficulty of handling
choice in a protocol.}]
Buyer: Seller!Request.
   ((Seller?Shipment.Seller!Payment) $\lor$ (Seller!Payment.Seller?Shipment))

Seller: Buyer?Request.
   ((Buyer!Shipment.Buyer?Payment) $\lor$ (Buyer?Payment.Buyer!Shipment))
\end{lstlisting}

Listing~\ref{Scribble:concurrency} shows how we might model
\pr~\ref{pr:flexible-purchase} in Scribble.  For the same reasons as
for Trace-C, the protocol in the listing is determined unrealizable by
Scribble.

\begin{lstlisting}[label={Scribble:concurrency},caption={Flexible purchase (\pr~\ref{pr:flexible-purchase}) in Scribble.}]
global protocol FlexiblePurchase (role Buyer, role Seller) {
 Request() from Buyer to Seller;
 choice at Buyer {
   Payment() from Buyer to Seller;
   Shipment() from Seller to Buyer;
 } or {
   Shipment() from Seller to Buyer; // not valid
   Payment() from Buyer to Seller;
 }
}
\end{lstlisting}

Some research branches of Scribble
\citep{demangeon:session-types:2015} have included a \emph{parallel}
operator, which however is absent from the main Scribble language and
implementation.  We hypothesize that a parallel operator would
manifest as a problematic nonlocal choice.  Our hypothesis is based on
the fact that Trace-C's $\land$ operator is in essence a parallel
operator and as we showed in the analysis of flexible purchase in
Trace-C, $\land$ manifests as a problematic nonlocal choice in the
projections.

\begin{figure}[htb!]
\centering
\begin{tikzpicture}[>=stealth']
\tikzstyle{every node}=[inner sep=0.4em]
\tikzstyle{every path}=[font=\small\sffamily]
\tikzstyle{m_label_base}=[draw=none,midway,fill=none,sloped,align=center,font=\small\sffamily]
\tikzstyle{m_label_up}=[m_label_base,above=-2pt]

\node[circle,draw] (s_0) at (0,0) {$s_0$};
\node[circle,draw] (s_1) at (6,0) {$s_1$};
\node[double,circle,draw] (s_2) at (13,0) {$s_2$};

\path [->] (s_0) edge node [m_label_up] {Buyer $\mo$ Seller: Request()} (s_1)
           (s_1) edge [loop above] node[xshift=1em] {Buyer $\mo$ Seller: Payment()[unbound(paid)]/bind(paid,T)} ()
                 edge [loop below] node[xshift=2em] {Seller $\mo$ Buyer: Shipment()[unbound(shipped)]/bind(shipped,T)} ()
                 edge node [m_label_up] {[bound(paid)$\land$bound(shipped)]} (s_2);
\node[anchor=east] at (s_0.west) {$\mathcal{P}\mathbf{:}$};

\end{tikzpicture}
\caption{\pname{FlexiblePurchase} in HAPN.}
\label{hapn:flexible-purchase}
\end{figure}
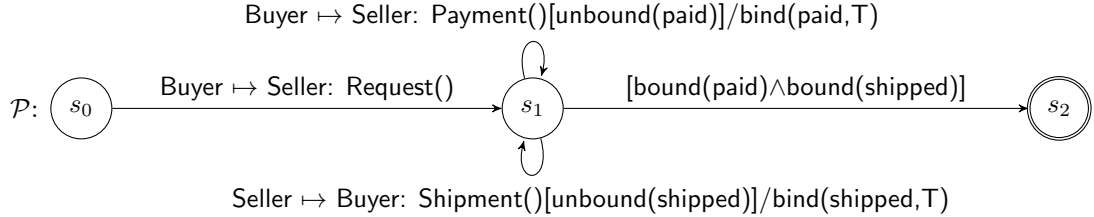

Figure~\ref{hapn:flexible-purchase}'s HAPN protocol captures only the
first two enactments of Figure~\ref{fig:concurrency}, not the
concurrent one because HAPN requires synchrony.

Listing~\ref{BSPL:flexible-purchase} gives a BSPL protocol.  It
supports the enactment in Figure~\ref{fig:crossing} because after
\rname{buyer} sends \mname{Request}, it has the information needed to
send \mname{Payment} and, upon receiving \mname{Request},
\rname{seller} has the information needed to send \mname{Shipment}.
The protocol also supports the enactments in
Figures~\ref{fig:shipment-first} and~\ref{fig:payment-first}.

\begin{lstlisting}[label={BSPL:flexible-purchase},caption={Flexible purchase (\pr~\ref{pr:flexible-purchase}) in BSPL.}]
Flexible Purchase {
 $\role$ Buyer, Seller
 $\param$ out ID key, out item, out shipped, out paid

 Buyer $\mo$ Seller: Request[out ID, out item]
 Seller $\mo$ Buyer: Shipment[in ID, in item, out shipped]
 Buyer $\mo$ Seller: Payment[in ID, in item, out paid]
}
\end{lstlisting}

\subsection{Extensibility}

Is the protocol language such that an agent may participate in
multiple, potentially unrelated protocols specified in it? If the
answer is yes, we refer to the language as \emph{extensible}.
Technically, extensibility means that an agent may interleave
observations of messages from several protocols and yet be compliant
with each of them.  If an agent may participate in only one protocol,
that is, observe messages from only one protocol, then the language
is nonextensible.

Extensibility is a practical necessity.  For example, an organization
as an agent may interact with other organizations using one protocol
but interact with its own members using another protocol.  Further,
nonextensibility would be an undue restriction on an agent's design
and therefore its principal's autonomy.

As basic a requirement as extensibility appears to be, we show below
that several languages are in fact not extensible.

\begin{example}[Pricing+Catalog]\label{pr:pricing+catalog}
  \rname{seller} and \rname{buyer} engage in the \pname{Pricing}
  protocol, by which a \rname{buyer} may obtain offers for requested
  items from \rname{seller}.  In addition, \rname{seller} engages with
  \rname{provider} via the \pname{Catalog} protocol to obtain
  information about the newest products.  Further, \rname{buyer} is
  unaware of \pname{Catalog} and \rname{provider} is unaware of
  \pname{Pricing}, indicating that these protocols are not composed
  into a single protocol.
\end{example}

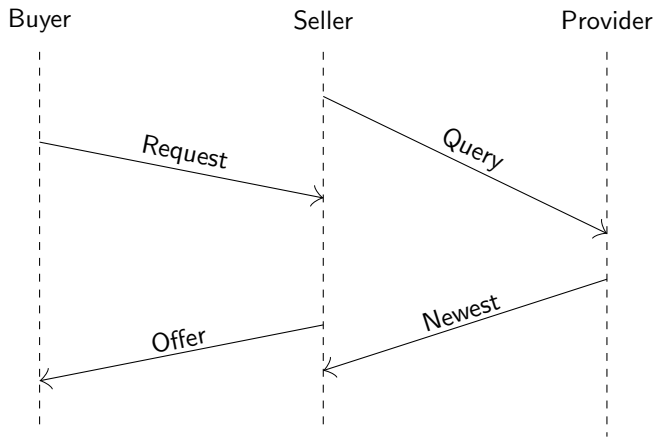
\begin{figure}[!htb]
\centering
\begin{tikzpicture}

  \tikzstyle{role}=[thin,draw,align=center,font=\small\sffamily,rectangle,anchor=center,minimum height=5ex,minimum width=6ex,inner sep=1]

  \tikzstyle{every text node part/.style}=[align=center]

  \tikzset{every node text/.style={node
      contents=\transform{#1}}}
\newcommand{\transform}[1]{\ensuremath{\mathsf{#1}}}

  \tikzstyle{m_label_base}=[draw=none,midway,fill=none,sloped,align=center,font=\small\sffamily]
  \tikzstyle{m_label_up}=[m_label_base,above=-2pt]
  \tikzstyle{m_label_down}=[m_label_base,below=-2pt]

 \tikzstyle{a_label}=[draw=none,sloped,fill=white,align=center,font=\small\scshape]

% \tikzstyle{message}=[->, >=stealth']
  \tikzstyle{emptybox}=[draw=none,minimum height=3ex]

\matrix () [ row sep=10,column sep=100] {
  \node[emptybox] (a) {};
  & \node[emptybox] (b) {};
  &  \node[emptybox] (c) {};\\
  \node (a-zero) {};
  & \node (b-zero) {};
  & \node (c-zero) {};\\
  \node (a-one) {};
  & \node (b-one) {};
  & \node (c-one) {};\\
  \node (a-two) {};
  & \node (b-two) {};
  & \node (c-two) {}; \\
  \node (a-three) {};
  & \node (b-three) {};
  & \node (c-three) {}; \\
  \node (a-four) {};
  & \node (b-four) {};
  & \node (c-four) {}; \\
  \node (a-five) {};
  & \node (b-five) {};
  & \node (c-five) {}; \\
  \node (a-six) {};
  & \node (b-six) {};
  & \node (c-six) {}; \\
  \node (a-seven) {};
  & \node (b-seven) {};
  & \node (c-seven) {}; \\
};

\node [role,draw=none,anchor=south] at (a) {Buyer};
\node [role,draw=none,anchor=south] at (b) {Seller};
\node [role,draw=none,anchor=south] at (c) {Provider};

\draw [dashed] (a.center)--(a-seven.center);
\draw [dashed] (b.center)--(b-seven.center);
\draw [dashed] (c.center)--(c-seven.south);

\draw [message] (a-one.north)--node [m_label_up] {Request} (b-two.center);
\draw [message] (b-zero.north)--node [m_label_up] {Query} (c-three.north);
\draw [message] (c-four.north)--node [m_label_up] {Newest} (b-six.north);
\draw [message] (b-five.north)--node [m_label_up] {Offer} (a-six.center);

\end{tikzpicture}
\caption{Extensibility means supporting enactments that interleave messages from two protocols, as shown.}
\label{fig:extensibility}
\end{figure}

Figure~\ref{fig:extensibility} shows an enactment of
\pr~\ref{pr:pricing+catalog} in which the messages of \pname{Pricing}
and \pname{Catalog} are interleaved.  In particular, notice that
\rname{seller} observes messages from both protocols.  There is
nothing in the enactment that tells us it should be deemed incorrect.
Such enactments would in fact be indicative of flexibility.  If a
protocol language were not extensible, then enactments such as the one
in Figure~\ref{fig:extensibility} would be deemed incorrect.

\begin{lstlisting}[caption={\pname{Catalog} and \pname{Pricing} in Trace-C.},label={Trace:catalog}]
\\Catalog  
Seller $\ar{Query}$ Provider; Provider $\ar{Newest}$ Seller

\\Pricing
Buyer $\ar{Request}$ Seller ; Seller $\ar{Offer}$ Buyer
\end{lstlisting}

Trace-C's semantics is in tension with extensibility.
Listings~\ref{Trace:catalog} gives the specifications of
\pname{Catalog} and \pname{Pricing} in Trace-C.
\review{B:fitness}\citet[p.~3]{Castagna+12:sessions} promote a
correctness criterion \emph{fitness} with respect to a protocol, which
says that an agent correctly implements (\emph{fits}) a protocol only
if it observes no message that is not in the protocol.  Castagna
{\etal} give the example of a protocol in which \rname{seller} may
send \rname{buyer} messages corresponding to \mname{price} and
\mname{descr} (of an item) and say that a seller that sent any message
other than \mname{price} and \mname{descr} would violate the protocol.
Naturally, it follows that if an agent observes messages from two or
more Trace-C protocols, then it correctly implements none of those
protocols.  In essence, fitness limits agents to implementing a single
protocol.

Returning to our example, Trace-C says that to correctly implement
\pname{Pricing}, \rname{seller} must not observe any message from
\pname{Catalog} and to correctly implement \pname{Catalog},
\rname{seller} must not observe any message from \pname{Pricing}.  An
agent that engages in both \pname{Pricing} and \pname{Catalog}
violates both.

A possible way to interleave interactions from two protocols in
Trace-C is to compose them into a single protocol, which an agent can
potentially fit.  Listing~\ref{Trace:pricing+catalog} shows a
composition of \pname{Pricing} and \pname{Catalog}.  The fitness of
\rname{seller} with respect to the composed protocol is thus a
possibility.  However, there are significant drawbacks to requiring
explicit composition just for the sake of fitness.  First, it would
create large unwieldy protocols with potentially unrelated
communications and introducing otherwise unrelated agents into the MAS
just because they are reachable from each other based on their
interactions with common agents.  Second, when protocols are large and
unwieldy, projections for agents and their implementations would
become correspondingly complicated.  Third, it would prevent any
organizational abstraction.  For example, the protocol by which an
organization trades with others will have to be composed with all the
protocols pertaining to interactions internal to the organization,
which would be undesirable.  Fourth, the composite protocol may turn
out to be unrealizable anyway.  Consider
Listing~\ref{Trace:pricing+catalog}, which gives a protocol,
\pname{Pricing+Catalog}, that composes \pname{Pricing} and
\pname{Catalog}.  This protocol is unrealizable.  The reason is that
the protocol sets up a problematic nonlocal choice (\rname{seller}
sends \mname{Query} or \rname{buyer} sends \mname{Request}).

\begin{lstlisting}[caption={\pname{Pricing+Catalog} in
    Trace-C.},label={Trace:pricing+catalog}]
// Pricing+Catalog
(Buyer $\ar{Request}$ Seller; Seller $\ar{Offer}$ Buyer) $\land $ (Seller $\ar{Query}$ Provider; Provider $\ar{Newest}$ Seller)
\end{lstlisting}

Based on the foregoing analysis, we conclude that Trace-C does not
support extensibility.  \review{B:ex-proj}Considering the
seller's projections for \mname{Pricing} and \mname{Catalog}, given in
Listing~\ref{Trace:pricing-catalog-proj}, helps see the point
intuitively.  If you consider the projection from \pname{Pricing}, the
only trace that satisfies projection is \mname{Request} followed by
\mname{Offer}.  In particular, no trace with \mname{Query} or
\mname{Newest} satisfies the projection from \pname{Pricing}.  By an
analogous argument, no trace with \mname{Request} or \mname{Offer}
would satisfy the projection from \pname{Catalog}.

\begin{lstlisting}[caption={Projections of \pname{Pricing} and \pname{Catalog} in
    Trace-C (and Trace-F) for Seller.},label={Trace:pricing-catalog-proj}]
//Seller's projection from Pricing
Seller: Buyer?Request ; Buyer!Offer

//Seller's projection from Catalog
Seller: Provider!Query ; Provider?Newest
\end{lstlisting}

\review{B:ex-others}Following an analogous line of reasoning, we can
see that Trace-F and Scribble do not support extensibility either:
\pname{Pricing} and \pname{Catalog} yield projections for
\rname{seller}, none of which entertains traces with events from the
other (in fact, the projections in
Listing~\ref{Trace:pricing-catalog-proj} double as projections in
Trace-F as well).  HAPN's state-machine semantics rules out
interleavings with other protocols.  Therefore, HAPN too does not
support extensibility.

\review{B:UoD}The underlying reason Trace-C, Trace-F, Scribble, and HAPN come up
short on extensibility is they all implicitly identify the universe of
discourse (the messages that agents may observe) with the messages
specified in the protocol.

BSPL supports extensibility.  In fact, BSPL refers to a universe of
discourse, which for an agent would include any messages it observes,
regardless of the protocols they feature in.  Any message schema is an
elementary protocol is BSPL and determining the correctness of an
observation of the message depends on the causality and integrity
constraints specified in the schema.  Correctness does not depend upon
the protocol in which a message schema occurs.
Listing~\ref{BSPL:pricing} and Listing~\ref{BSPL:catalog} specify
\pname{Pricing} and \pname{Catalog}, respectively, in BSPL.  The role
\rname{seller} features in both protocols and interleaves observations
of messages from both protocols.

\begin{lstlisting}[caption={Pricing in BSPL.},label={BSPL:pricing}]
Pricing {
 $\role$ Buyer, Seller
 $\param$ out ID $\msf{key}$, out item, out price

 Buyer $\mo$ Seller: Request[out ID, out item]
 Seller $\mo$ Buyer: Offer[in ID, out price]
}
\end{lstlisting}

\begin{lstlisting}[caption={\pname{Catalog} in BSPL.},label={BSPL:catalog}]
Catalog {
 $\role$ Seller, Provider
 $\param$ out qID key, out req, out products

 Seller $\mo$ Provider: Query[out qID, out req]
 Provider $\mo$ Seller: Newest[in qID, in req, out products]
}
\end{lstlisting}

\section{Information Modeling}\review{B:name}
\label{sec:information}

Does a language enable capturing the information model underlying the
interactions in a MAS?  Further, does the information model captured
in a protocol specification enable computing social meaning correctly?

It is the information conveyed via messaging that enables coordination
in a MAS.
\review{B:mot2}
Such information has a model, starting with constraints such as would
be captured in a message schema, e.g., to capture message instances,
and extending to constraints between message schemas in a protocol
specification, e.g., to capture correlation and integrity in protocol
enactments, which may be viewed as groups of message instances.
Further, due to Constraint~\ref{req:social} states, since a
meaning-level event is computed as a view over one or more protocol
events, the soundness of
\review{B:social}
information at the meaning-level relies on the soundness of
information generated in protocol-level events.

We elaborate below on protocol instances and the integrity of
information as a lead up to social meaning.

\subsection{Protocol Instances}

A practical requirement is that a protocol, being a pattern of
communication, may be instantiated several times.  We refer to each
instantiation as a protocol instance that would be comprised of some
appropriately correlated messages.  How well do protocol languages
support protocol instances?  Consider \pr~\ref{pr:concurrent-pricing}.

\begin{example}[Concurrent Pricing]\label{pr:concurrent-pricing}
\review{B:usecase}
  A buyer and seller may engage in several, possibly concurrent
  engagements, in each of which a buyer sends a request for some item
  and the seller responds with an offer.
\end{example}

\begin{figure}[htb!]
  \centering
\begin{subfigure}[t]{0.45\columnwidth}
  \centering
\begin{tikzpicture}

  \tikzstyle{role}=[thin,draw,align=center,font=\small\sffamily,rectangle,anchor=center,minimum height=5ex,minimum width=6ex,inner sep=1]

  \tikzstyle{every text node part/.style}=[align=center]

  \tikzset{every node text/.style={node
      contents=\transform{#1}}}
\newcommand{\transform}[1]{\ensuremath{\mathsf{#1}}}

  \tikzstyle{m_label_base}=[draw=none,midway,fill=none,sloped,align=center,font=\small\sffamily]
  \tikzstyle{m_label_up}=[m_label_base,above=-2pt]
  \tikzstyle{m_label_down}=[m_label_base,below=-2pt]

 \tikzstyle{a_label}=[draw=none,sloped,fill=white,align=center,font=\small\scshape]

  \tikzstyle{emptybox}=[draw=none,minimum height=3ex]

\matrix () [row sep=10,column sep=130] {
  \node[emptybox] (a) {};
  &  \node[emptybox] (b) {};\\
  \node (a-zero) {};
  & \node (b-zero) {}; \\
  \node (a-one) {};
  & \node (b-one) {}; \\
  \node (a-two) {};
  & \node (b-two) {}; \\
  \node (a-three) {};
  & \node (b-three) {}; \\[1]
  \node (a-four) {};
  & \node (b-four) {}; \\
  \node (a-five) {};
  & \node (b-five) {}; \\
  \node (a-six) {};
  & \node (b-six) {}; \\
  \node[emptybox] (ae) {};
  &  \node[emptybox] (be) {};\\
};

\node [role,draw=none,anchor=south] at (a) {Buyer};
\node [role,draw=none,anchor=south] at (b) {Seller};

\draw [dashed] (a.center)--(ae.center);
\draw [dashed] (b.center)--(be.center);

\draw [message] (a-zero.north)--node [m_label_up] {Request(1, fig)} (b-one.north);

\draw [message] (b-two.north)--node [m_label_up] {Offer(1, \$5)} (a-three.north);

\draw [message] (a-four.north)--node [m_label_up,pos=0.7] {Request(2, jam)} (b-five.north);

\draw [message] (b-five.south)--node [m_label_up,pos=0.7] {Offer(2, \$6)} (a-six.south);

%\node[font=\bfseries] at ($(ae) !  0.5 !  (be)$)  {\vdots};

\end{tikzpicture}
\caption{Serial: \rname{buyer} sends the second \mname{Request} after receiving an \mname{Offer} for the first.}
\label{fig:sequential}
\end{subfigure}
\hfill
\begin{subfigure}[t]{0.45\columnwidth}
  \centering
\begin{tikzpicture}

  \tikzstyle{role}=[thin,draw,align=center,font=\small\sffamily,rectangle,anchor=center,minimum height=5ex,minimum width=6ex,inner sep=1]

  \tikzstyle{every text node part/.style}=[align=center]

  \tikzset{every node text/.style={node
      contents=\transform{#1}}}
\newcommand{\transform}[1]{\ensuremath{\mathsf{#1}}}

  \tikzstyle{m_label_base}=[draw=none,midway,fill=none,sloped,align=center,font=\small\sffamily]
  \tikzstyle{m_label_up}=[m_label_base,above=-2pt]
  \tikzstyle{m_label_down}=[m_label_base,below=-2pt]

 \tikzstyle{a_label}=[draw=none,sloped,fill=white,align=center,font=\small\scshape]

% \tikzstyle{message}=[->, >=stealth']
  \tikzstyle{emptybox}=[draw=none,minimum height=3ex]

\matrix () [row sep=10,column sep=130] {
  \node[emptybox] (a) {};
  &  \node[emptybox] (b) {};\\
  \node (a-zero) {};
  & \node (b-zero) {}; \\
  \node (a-one) {};
  & \node (b-one) {}; \\
  \node (a-two) {};
  & \node (b-two) {}; \\
  \node (a-three) {};
  & \node (b-three) {}; \\[1]
  \node (a-four) {};
  & \node (b-four) {}; \\
  \node (a-five) {};
  & \node (b-five) {}; \\
  \node (a-six) {};
  & \node (b-six) {}; \\
  \node[emptybox] (ae) {};
  &  \node[emptybox] (be) {};\\
};
\node [role,draw=none,anchor=south] at (a) {Buyer};
\node [role,draw=none,anchor=south] at (b) {Seller};

\draw [dashed] (a.center)--(ae.center);
\draw [dashed] (b.center)--(be.center);

\draw [message] (a-zero.north)--node [m_label_up] {Request(1, fig)} (b-one.south);
\draw [message] (a-two.north)--node [m_label_up] {Request(2, jam)} (b-three.south);
\draw [message] (b-four.north)--node [m_label_up] {Offer(2, \$6)} (a-five.south);
\draw [message] (b-five.north)--node [m_label_up] {Offer(1, \$5)} (a-six.south);

%\node[font=\bfseries] at ($(ae) !  0.5 !  (be)$)  {\vdots};

\end{tikzpicture}
\caption{Second first: \rname{buyer} sends two \mname{Request}s.  They
  arrive at the \rname{seller} in the same order as they were sent.
  \rname{seller} responds to each \mname{Request} with an
  \mname{Offer} but in the reverse order.}
\label{fig:rfq-replies}
\end{subfigure}
\hfill
\begin{subfigure}[t]{0.45\columnwidth}
  \centering
\begin{tikzpicture}

  \tikzstyle{role}=[thin,draw,align=center,font=\small\sffamily,rectangle,anchor=center,minimum height=5ex,minimum width=6ex,inner sep=1]

  \tikzstyle{every text node part/.style}=[align=center]

  \tikzset{every node text/.style={node
      contents=\transform{#1}}}
  \newcommand{\transform}[1]{\ensuremath{\mathsf{#1}}}

    \tikzstyle{m_label_base}=[draw=none,midway,fill=none,sloped,align=center,font=\small\sffamily]
  \tikzstyle{m_label_up}=[m_label_base,above=-2pt]
  \tikzstyle{m_label_down}=[m_label_base,below=-2pt]

 \tikzstyle{a_label}=[draw=none,sloped,fill=white,align=center,font=\small\scshape]

% \tikzstyle{message}=[->, >=stealth']
  \tikzstyle{emptybox}=[draw=none,minimum height=3ex]

\matrix () [row sep=10,column sep=130] {
  \node[emptybox] (a) {};
  &  \node[emptybox] (b) {};\\
  \node (a-zero) {};
  & \node (b-zero) {}; \\
  \node (a-one) {};
  & \node (b-one) {}; \\
  \node (a-two) {};
  & \node (b-two) {}; \\
  \node (a-three) {};
  & \node (b-three) {}; \\[1]
  \node (a-four) {};
  & \node (b-four) {}; \\
  \node (a-five) {};
  & \node (b-five) {}; \\
  \node (a-six) {};
  & \node (b-six) {}; \\
  \node[emptybox] (ae) {};
  &  \node[emptybox] (be) {};\\
};
\node [role,draw=none,anchor=south] at (a) {Buyer};
\node [role,draw=none,anchor=south] at (b) {Seller};

\draw [dashed] (a.center)--(ae.center);
\draw [dashed] (b.center)--(be.center);

\draw [message] (a-zero.north)--node [m_label_up] {Request(1, fig)} (b-one.north);

\draw [message] (b-two.south)--node [m_label_up,pos=0.25] {Offer(1, \$5)} (a-four.south);

\draw [message] (a-two.south)--node [m_label_up,,pos=0.25] {Request(2, jam)} (b-four.south);

\draw [message] (b-five.south)--node [m_label_up] {Offer(2, \$6)} (a-six.south);

%\node[font=\bfseries] at ($(ae) !  0.5 !  (be)$)  {\vdots};

\end{tikzpicture}
\caption{Concurrent: \rname{buyer} sends a second \mname{Request}; concurrently, \rname{seller} responds to the first with an \mname{Offer}.  The messages cross in transit.}
\label{fig:concurrent}
\end{subfigure}
\hfill
\begin{subfigure}[t]{0.45\columnwidth}
  \centering
\begin{tikzpicture}

  \tikzstyle{role}=[thin,draw,align=center,font=\small\sffamily,rectangle,anchor=center,minimum height=5ex,minimum width=6ex,inner sep=1]

  \tikzstyle{every text node part/.style}=[align=center]

  \tikzset{every node text/.style={node
      contents=\transform{#1}}}
  \newcommand{\transform}[1]{\ensuremath{\mathsf{#1}}}

    \tikzstyle{m_label_base}=[draw=none,midway,fill=none,sloped,align=center,font=\small\sffamily]
  \tikzstyle{m_label_up}=[m_label_base,above=-2pt]
  \tikzstyle{m_label_down}=[m_label_base,below=-2pt]

 \tikzstyle{a_label}=[draw=none,sloped,fill=white,align=center,font=\small\scshape]

% \tikzstyle{message}=[->, >=stealth']
  \tikzstyle{emptybox}=[draw=none,minimum height=3ex]

\matrix () [row sep=10,column sep=130] {
  \node[emptybox] (a) {};
  &  \node[emptybox] (b) {};\\
  \node (a-zero) {};
  & \node (b-zero) {}; \\
  \node (a-one) {};
  & \node (b-one) {}; \\
  \node (a-two) {};
  & \node (b-two) {}; \\
  \node (a-three) {};
  & \node (b-three) {}; \\[1]
  \node (a-four) {};
  & \node (b-four) {}; \\
  \node (a-five) {};
  & \node (b-five) {}; \\
  \node (a-six) {};
  & \node (b-six) {}; \\
  \node[emptybox] (ae) {};
  &  \node[emptybox] (be) {};\\
};
\node [role,draw=none,anchor=south] at (a) {Buyer};
\node [role,draw=none,anchor=south] at (b) {Seller};

\draw [dashed] (a.center)--(ae.center);
\draw [dashed] (b.center)--(be.center);

\draw [message] (a-zero.north)--node [m_label_down] {Request(1, fig)} (b-four.north);

\draw [message] (a-one.north)--node [m_label_up,pos=0.75] {Request(2, jam)} (b-two.north);
\draw [message] (b-three.north)--node [m_label_down,pos=0.5] {Offer(2, \$6)} (a-four.south);
\draw [message] (b-five.north)--node [m_label_up] {Offer(1, \$5)} (a-six.south);

%\node[font=\bfseries] at ($(ae) !  0.5 !  (be)$)  {\vdots};

\end{tikzpicture}
\caption{Out of order: \rname{buyer} sends two \mname{Request}s, which cross in transit.  \rname{seller} responds to the \mname{Request}s with \mname{Offer}s in the order they arrive.}
\label{fig:out-of-order}
\end{subfigure}
\caption{Four possible enactments of \pr~\ref{pr:concurrent-pricing}, in
  each of which buyer and seller engage in two instances of
  \pname{Pricing}.  Messages with identifier \texttt{1} belong to one
  instance and messages with identifier \texttt{2} to the other
  instance.}
\label{fig:instances}
\end{figure}
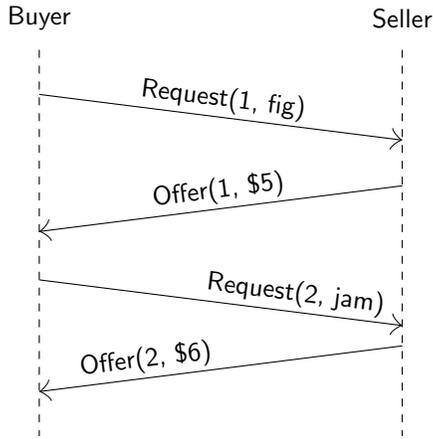
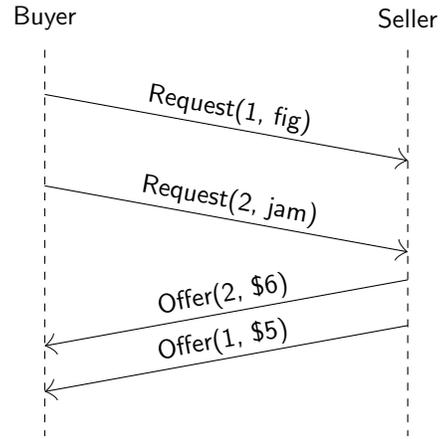
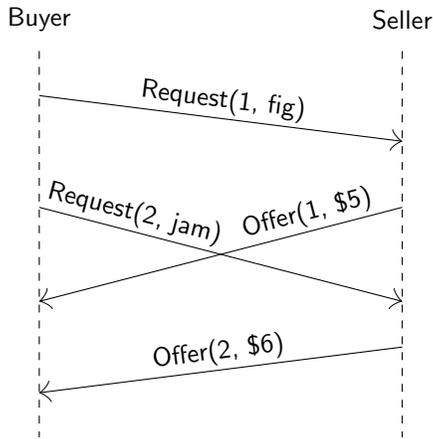
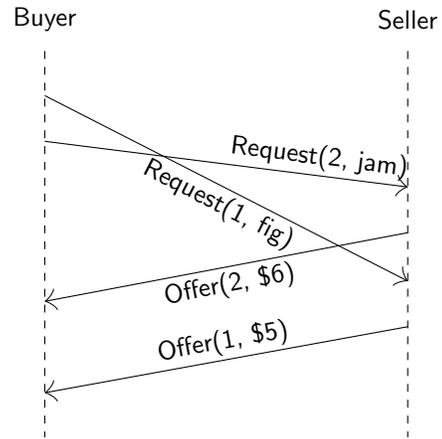

Figure~\ref{fig:instances} illustrates some enactments involving two
instances of the pattern in \pr~\ref{pr:concurrent-pricing}.  The messages
contain identifiers (\val{1} and \val{2}) to help distinguish the
instances from each other and to correlate messages within an
instance.

Listing~\ref{Scribble:concurrent-pricing} gives a candidate Scribble protocol
capturing \pr~\ref{pr:concurrent-pricing}.  Notice the recursion, which crucially
enables \rname{buyer} and \rname{seller} to engage repeatedly in the
pricing pattern of \pr~\ref{pr:concurrent-pricing}.  However, each pricing
engagement must happen in its entirety before another can start; that
is, the protocol does not allow interleaving of multiple pricing
engagements.  Thus, although the protocol supports the enactment of
Figure~\ref{fig:sequential}, it excludes the enactments of
Figures~\ref{fig:rfq-replies} and~\ref{fig:concurrent}.  The enactment
of Figure~\ref{fig:out-of-order} is also excluded but for a different
reason: it violates FIFO, which is a requirement for Scribble.

\begin{lstlisting}[caption={Concurrent Pricing (\pr~\ref{pr:concurrent-pricing}) in Scribble.},label={Scribble:concurrent-pricing}]
global protocol Pricing(role Buyer, role Seller) {
  Request(ID : String, item : String) from Buyer to Seller;
  Offer(ID, price : String) from Seller to Buyer;
  do Pricing(Buyer, Seller);
}
\end{lstlisting}

Listing~\ref{Trace-C:concurrent-pricing} gives a Trace-C protocol for
\pr~\ref{pr:concurrent-pricing}.  Here, the $^*$ means that the enclosed pattern
may be repeated zero or more times.  As for Scribble, each iteration
must complete before another can begin, thereby excluding the
enactments of Figures~\ref{fig:rfq-replies} and~\ref{fig:concurrent}
due to the semantics of the language and excluding
Figure~\ref{fig:out-of-order} due to the FIFO requirement.

\begin{lstlisting}[caption={Concurrent Pricing (\pr~\ref{pr:concurrent-pricing}) in Trace-C.},label={Trace-C:concurrent-pricing}]
  (Buyer $\ar{Request(ID, item)}$ Seller ; Seller $\ar{Offer(ID, price)}$ Buyer)$^*$
\end{lstlisting}

Listing~\ref{Trace-F:concurrent-pricing} gives a recursive Trace-F protocol for
\pr~\ref{pr:concurrent-pricing}.  As for Scribble and Trace-C, each
iteration must complete before another can begin, thereby excluding
the enactments of Figures~\ref{fig:rfq-replies}
and~\ref{fig:concurrent} due to the semantics of the language.  Under
unordered asynchrony, the protocol is unrealizable because the
enactment in Figure~\ref{fig:out-of-order} may occur but the protocol
cannot handle it.  Under FIFO asynchrony,
Figure~\ref{fig:out-of-order} is excluded, just as for Scribble and
Trace-C.

\begin{lstlisting}[caption={Concurrent Pricing (\pr~\ref{pr:concurrent-pricing}) in Trace-F.},label={Trace-F:concurrent-pricing}]
  P = Buyer $\ar{Request(ID, item)}$ Seller ; Seller $\ar{Offer(ID, price)}$ Buyer ; P
\end{lstlisting}

With the aim of supporting the enactment in
Figure~\ref{fig:concurrent}, Listing~\ref{Trace-F:interleaved-pricing}
gives an alternative Trace-F specification.  This specification has
two problems.
\review{B:parallel}
One, it would support at most two instances.  Two, and
in any case, it is unrealizable.  This is because it would manifest in
a problematic nonlocal choice in the projections for \rname{buyer} and
\rname{seller}: after sending the first \mname{Request}, \rname{buyer}
can either send the second \mname{Request} or receive \mname{Offer},
and after receiving the first \mname{Request}, \rname{seller} can
either send \mname{Offer} or receive the second \mname{Request}.

\begin{lstlisting}[caption={Another attempt at Concurrent Pricing (\pr~\ref{pr:concurrent-pricing}) in Trace-F.},label={Trace-F:interleaved-pricing}]
  (Buyer $\ar{Request(ID, item)}$ Seller ; Seller $\ar{Offer(ID, price)}$ Buyer) $\land$ (Buyer $\ar{Request(ID, item)}$ Seller ; Seller $\ar{Offer(ID, price)}$ Buyer)
\end{lstlisting}

Listing~\ref{Trace-F:interleaved-recursive-pricing} (unclear if it is
a legal Trace-F specification) improves upon
Listing~\ref{Trace-F:interleaved-pricing} by allowing an unbounded
number of instances; however, it remains unrealizable for the same
reason as Listing~\ref{Trace-F:interleaved-pricing}.

\begin{lstlisting}[caption={Concurrent Recursive Pricing (\pr~\ref{pr:concurrent-pricing}) in Trace-F.},label={Trace-F:interleaved-recursive-pricing}]
  P = Buyer $\ar{Request(ID, item)}$ Seller ; Seller $\ar{Offer(ID, price)}$ Buyer $\land$ P
\end{lstlisting}

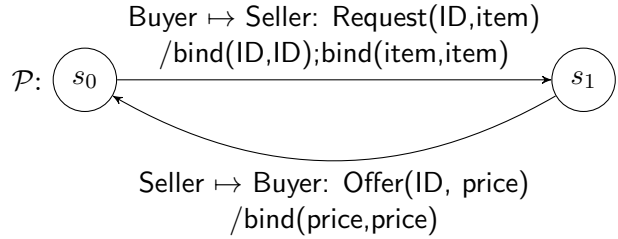
\begin{figure}[htb!]
\centering
\begin{tikzpicture}[>=stealth']
\graph [nodes={circle,inner sep=0.4em,draw}, math nodes,
        edges={align=center,font=\sffamily},
        grow right sep=15em]
{
  s_0/"s_0"-> ["Buyer $\mo$ Seller: Request(\paraname{ID},\paraname{item})\\/bind(ID,\paraname{ID});bind(item,\paraname{item})"] s_1;
  s_1-> ["Seller $\mo$ Buyer: Offer(\paraname{ID}, \paraname{price})\\/bind(price,\paraname{price})", bend left] s_0;
};
\node[anchor=east] at (s_0.west) {$\mathcal{P}\mathbf{:}$};
\end{tikzpicture}
\caption{Concurrent Pricing in HAPN.}
\label{hapn:concurrent-pricing}
\end{figure}

Figure~\ref{hapn:concurrent-pricing} specifies the protocol in HAPN.
As for Scribble, Trace-C, and the Trace-F protocol in
Listing~\ref{Trace-F:concurrent-pricing}, each iteration must complete
before another can begin, thereby excluding the enactments of
Figures~\ref{fig:rfq-replies} and~\ref{fig:concurrent}.  Notice that
the requirement of synchrony eliminates Figures~\ref{fig:concurrent}
and~\ref{fig:out-of-order}.  That is, there are two strikes against
Figure~\ref{fig:concurrent}.

Because Scribble, Trace-C, and Trace-F support one of the four
enactments in Figure~\ref{fig:instances}, we conclude that they
partially support instances.  By contrast, BSPL fully supports the
specification of instances via \emph{key} parameters of protocols.
Listing~\ref{BSPL:pricing} in fact gives the BSPL protocol that
supports all the enactments in Figure~\ref{fig:instances}.

\review{B:instances}To help understand why the protocol in
Listing~\ref{BSPL:pricing} supports each of the enactments in
Figure~\ref{fig:instances}, the following observations suffice.
First, \rname{buyer} can send a \mname{Request} at any point because
it can generate bindings for both parameters of \mname{Request},
namely, \paraname{ID} and \paraname{item}.  However, as \mname{ID} is
key, no two \mname{Request}s may contain the same binding for
\paraname{ID}.  Second, \rname{seller} may send \mname{Offer}s only
for those \paraname{ID}s whose bindings it knows from prior
communications---here, from \mname{Request} messages it has received.
Further, once \rname{seller} knows a binding for \paraname{ID}, it can
send (depending on whether its reasoner determines that it should
send) an \mname{Offer} with that binding of \paraname{ID} at any point
since the protocol allows it to generate a binding for the only other
parameter in \mname{Offer}, namely, \paraname{price}.

\subsection{Integrity}

\review{B:integrity}
Building upon instances, \emph{integrity} means that information in a
protocol instance must not be inconsistent.  For example, in any
instance of \pname{Purchase}, \paraname{item} must have a unique
binding; it cannot be \textsl{fig} in \mname{Request} and \textsl{jam}
in \mname{Offer}. 

Scribble does not support integrity.  In Scribble, the
message names and the data types of the information carried in a
message matter; however, the information carried in the message does
not matter.  The motivating example of a Scribble travel booking
protocol \citep[p.~8]{Yoshida+13:Scribble} is illuminating in this
regard.  Listing~\ref{Scribble:no-info} reproduces the relevant parts
of that protocol.

\begin{lstlisting}[caption={Relevant parts of a Scribble travel booking protocol \citep{Yoshida+13:Scribble}.},label={Scribble:no-info}]
global protocol BookJourney(role Customer as C, role Agency as A, role Service as S) {
  ...
 query(journey: String) from C to A;
 price(Int) from A to C;
  ...
}
\end{lstlisting}

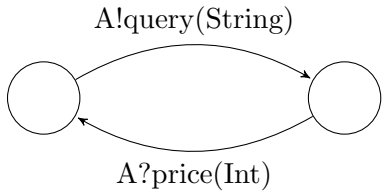
\begin{figure}[htb!]
\centering
\begin{tikzpicture}[>=stealth',shorten >=1pt,auto,node distance=4cm]
  \node[state] (q1)      {};
  \node[state] (q2) [right of=q1]  {};
  \path[->]          (q1)  edge [bend left] node {A!query(String)} (q2);
  \path[->]          (q2)  edge  [bend left] node {A?price(Int)} (q1);
\end{tikzpicture}
\caption{Part of \rname{customer}'s FSM, derived from Listing~\ref{Scribble:no-info}.  Here, A is the \rname{agency} role with which \rname{customer} interacts.}
\label{fig:Scribble-FSM}
\end{figure}

Figure~\ref{fig:Scribble-FSM} partially reproduces the
\rname{customer}'s finite state machine (FSM) that is extracted from
the above protocol and that serves as the basis for compliance
checking in the agent: a deviation from the FSM is a violation of the
protocol \citep[p.~10]{Yoshida+13:Scribble}.  Notice that the
parameter \paraname{journey} is absent; all that matters is that
\rname{customer} sends a \textit{String} to \rname{agency}.  To drive
home the point about the lack of information modeling in Scribble, we
refer the reader to the implementation of the \rname{customer} agent
\citep[p.~11]{Yoshida+13:Scribble}, which does not mention
\paraname{journey}.

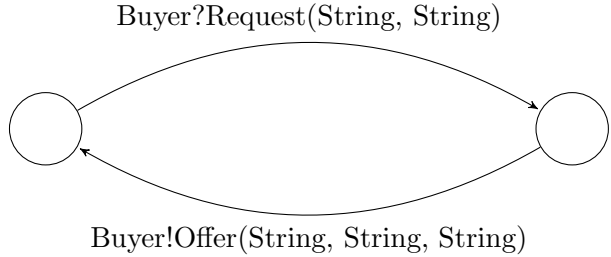
\begin{figure}[htb!]
\centering
\begin{tikzpicture}[>=stealth',shorten >=1pt,auto,node distance=7cm]
  \node[state] (q1)      {};
  \node[state] (q2) [right of=q1]  {};
%  \node[state] (q3) [right of=q2]  {};
  \path[->]          (q1)  edge[bend left] node {Buyer?Request(String, String)} (q2);
  \path[->]          (q2)  edge[bend left]  node {Buyer!Offer(String, String, String)} (q1);
\end{tikzpicture}
\caption{The \rname{seller}'s FSM, derived from Listing~\ref{Scribble:pricing-alt}}
\label{fig:Scribble-alt-FSM}
\end{figure}

Returning to the domain of our running examples, consider the Scribble
protocol in Listing~\ref{Scribble:pricing-alt}, a variant of
Listing~\ref{Scribble:concurrent-pricing}.  Specifically, in the
\pname{Alt-Pricing} protocol, \mname{Offer} additionally includes the
parameter \paraname{item}.  Figure~\ref{fig:Scribble-alt-FSM} gives
the \rname{seller}'s FSM.  Notice again that the parameter names are
absent; what matters are the data types.  This machine would determine
to be legal even those protocol enactments that violate integrity,
e.g., where \mname{Request} contains (the item) \val{fig} but
\mname{Offer} contains (the \paraname{item} binding) \val{jam}.

\begin{lstlisting}[caption={Alternative Pricing in Scribble.},label={Scribble:pricing-alt}]
global protocol Alt-Pricing(role Buyer, role Seller) {
  Request(ID:String, item:String) from Buyer to Seller;
  Offer(ID, item, price:String) from Seller to Buyer;
  do Pricing(Buyer, Seller);
}
\end{lstlisting}

% Because Scribble does not semantically entertain instances,
% repetitions as allowed by the recursion in \pname{Alt-Pricing}
% (Listing~\ref{Scribble:pricing-alt}) pose a challenge to introducing
% integrity in the language.  For example, we might imagine a simple
% method such as the following: if a parameter gets bound in some
% message, then we check that any subsequent message that features the
% parameter has the same binding for the parameter.  Thus, for example,
% if \mname{Request} binds \paraname{ID} and \paraname{item} to \val{1}
% and \val{fig}, respectively, then \mname{Offer} must also have the
% same bindings for those parameters.  What happens, though, in a
% repetition?  Consider a subsequent \mname{Request}.  Must it have the
% same bindings for those parameters?  Clearly, that would be
% unreasonable since \rname{buyer} may want a different \paraname{item}.
% Must the bindings be different?  That would be unreasonable too since
% a \rname{buyer} may want the same \paraname{item}.

Like Scribble, Trace-C does not support integrity either.  Like
Scribble, messages are opaque in Trace-C: the message names matter but
their contents do not, as Listing~\ref{Trace-C:Purchase} illustrates.

HAPN partially guarantees information integrity.  Once a variable is
bound in an enactment, as \paraname{item} is at state $s_1$ in
Figure~\ref{hapn:concurrent-pricing}, any message attempting to use
that variable with a different value is illegal.  HAPN's ``unbind''
operation does not compromise integrity under its assumptions of
synchronicity and shared state; all agents would simultaneously see
any updates, and would never have inconsistent bindings for any
variable.  However, HAPN does not ensure integrity when implementing
the protocol in a decentralized asynchronous environment, expecting
the designer to check for realizability and to make corrections to any
problems arising.

BSPL supports integrity, as described in Section~\ref{sec:bspl}.

Some authors of Trace-F have investigated a language closely related
to Trace-F that supports parameters \citep{ancona:parametric:2017}.
This language (which we dub Trace-Parameters), like HAPN, partially
captures integrity by supporting parameter bindings.
Listing~\ref{Trace-F:integrity} (reproduced from
\citep{mascardi:personal:2019}) gives a protocol in Trace-Parameters.
The expression on the right-hand side of the $=$ may be read as a
binding for a parameter (here, \paraname{ID}) followed by the scope
over which that binding holds.  That is, \paraname{ID} must have the
same fresh binding throughout the specified scope.  The fresh binding
mechanism is not sufficiently expressive to capture integrity, since
integrity needs identifiers (as BSPL supports via key parameters).
Specifically, the protocol in Listing~\ref{Trace-F:integrity} allows
two \mname{Request} messages to be sent with the same binding for
\paraname{ID}.  Not only that, it allows two \mname{Request} messages
with the same binding of \paraname{ID} to have different bindings for
\paraname{item}.

\begin{lstlisting}[caption={A protocol in Trace-Parameters that requires the same binding for \paraname{ID} throughout in every recursion instance.},label={Trace-F:integrity}]
  T = ID .  Buyer $\ar{Request(ID, item)}$ Seller ; Seller $\ar{Offer(ID, price)}$ Buyer ; T
\end{lstlisting}

\noindent \emph{Additional Remarks.}  \review{B:conversation} In the
absence of protocol language support for protocol instances and
correlation, approaches may resort to using the syntactic notion of a
\emph{conversation identifier} to identify messages belonging to the
same protocol instance.  FIPA ACL (implemented in the
JADE platform \citep{bellifemine:jade:2007}), adopts such an approach,
\review{C:F-J}
supporting the tagging of messages with conversation identifiers.  
Relying upon conversation identifiers though has significant
drawbacks.

\begin{itemize}
\item The burden of checking that conversation identifiers were being
  used appropriately would fall upon agent developers, for example, to
  check that no two \mname{Request}s are sent with the same
  identifier.
\item More importantly, since conversation identifiers represent an
  extra-protocol mechanism, using them would be a violation of
  Constraint~\ref{req:fullness} and would couple agents at the level
  of their implementations.
\item The required correlation may in any case go beyond whatever a
  single identifier supports.  For example, a single identifier would
  be ineffective for describing one-to-many relationships, which
  require composite identifiers.
\end{itemize}

\subsection{Social Meaning}
\label{sec:social-meaning}

In general, a specification of social meaning specifies the lifecycle
of meaning instances, as exemplified by recent work on commitments
\citep{chopra:cupid:2015}.  
\review{B:social}
For example, the Deliver-Payment commitment specification in
\pr~\ref{pr:commitment} specifies the lifecycle of Deliver-Payment
commitment instances.  Where each of the instances differ would be in
their information content, as \pr~\ref{pr:meaning-instance}
illustrates.

\begin{example}[Commitment instances]\label{pr:meaning-instance}
  \rname{buyer} and \rname{seller} repeatedly engage in purchase
  (\pr~\ref{pr:purchase}).  A distinct instance of the
  Deliver-Payment commitment (\pr~\ref{pr:commitment}) is created
  for every \mname{Accept}.  Several commitment instances may exist at
  any moment.
\end{example}

The information content of a commitment instance would need to obey
certain soundness constraints.  Each commitment instance referred to
in \pr~\ref{pr:meaning-instance} would involve a specific
\paraname{item} binding and a specific \paraname{price} binding.  And
in fact, those bindings must be \emph{immutable} over the entire
lifecycle of the instance.  In particular, it cannot be that the
commitment instance is \emph{created} with \paraname{item} binding
\val{fig} and \emph{discharged} with \paraname{item} binding
\val{jam}; such an enactment would clearly be unsound.  Notably,
reasoning about the state of a commitment instance involves reasoning
about several correlated messages that satisfy integrity---e.g., a
message to create the instance and a message to discharge it.

Given that meaning-level information is derived from (is a \emph{view}
over) underlying protocol events (Constraint~\ref{req:social}),
identifiers for commitment instances and the necessary correlation and
integrity already be present in protocol enactments.  For each
instance of the Deliver-Payment commitment, an instance of
\pname{Purchase} must generate the relevant information in a sound
manner: an identifier to identify the commitment instance and the
bindings for \paraname{item} and \paraname{price}.  This means that a
commitment instance created by a particular \mname{Accept} (belonging
to a particular protocol instance) may be discharged only by a
properly correlated instance of \mname{Payment} (belonging to that
same protocol instance).  And integrity of the protocol instance would
ensure that the bindings are immutable over the lifecycle of the
commitment instance.

It was common in early work on commitments to introduce arbitrary
(syntactic) identifiers for them disconnected from the underlying
information.  As \citet{AAMAS-Alignment-09} show, such schemes are
incompatible with commitment reasoning.

In a nutshell, the soundneed of computations at the commitment level
and more generally meaning level imposes requirements relating to
protocol instances and their integrity on the protocol language.
Scribble, Trace-C, Trace-F, and HAPN at best partially support
instances and integrity and therefore fall short on this criterion.
BSPL supports both instances and integrity and therefore better
supports social meaning.  Indeed, \citet[p.~498]{AAMAS-BSPL-11}
discusses social meaning and other work has applied BSPL toward
commitment consistency in decentralized settings
\citep{king:tosca:2017}.  Listing~\ref{cupid:del-pay} specifies the
Deliver-Payment commitment in Cupid \citep{chopra:cupid:2015}.  It
says that each instance of this specification is from \rname{Buyer} to
\rname{Seller}; it is created upon an \mname{Accept}; detached upon a
\mname{Deliver} that happens within three days of \mname{Accept}; and
discharged upon a \mname{Payment} that happens within three days of
\mname{Payment}.  Notably, the events in the commitment specification
refer to observations of messages of BSPL \pname{Purchase} protocol
given in Listing~\ref{BSPL:Purchase}.  Since BSPL supports protocol
instances and their integrity, as a protocol instance progresses, the
commitment instance too progresses soundly.  

\begin{lstlisting}[label={cupid:del-pay},caption={Deliver-Payment commitment in Cupid.}]
$\commit$ Deliver-Payment Buyer $\toward$ Seller
 $\create$ Accept
 $\detach$ Deliver [, Accept + 3]
 $\discharge$ Payment [, Deliver + 3]
\end{lstlisting}

\section{Operational Environment}
\label{sec:operational}

How strong are the assumptions that a language makes of the
operational environment of a multiagent system? The properties a
language requires of a communication infrastructure are assumptions
about the agent's operational environment.  The stronger the
assumptions a language makes, the more restrictive and less practical
the language.

We split this criterion into two: whether a language can work over
asynchronous infrastructures and if so, whether it can work with
unordered message delivery.

\subsection{Asynchronous Communication}

Assuming synchronous communication would be so strong an assumption as
to make the language impractical for MAS.  Asynchronous communication,
on the other hand, promotes loose coupling between agents and is also
practical in that it matches real-world constraints.  Scribble, Trace,
and BSPL accommodate asynchrony but HAPN
\citep[p.~61]{Winikoff+18:HAPN} does not.

\subsection{Unordered Communication}
\label{sec:unordered}

Asynchronous communication though comes in many flavors.  Asynchronous
communication with some kind of ordered delivery, for example,
\emph{pairwise FIFO} (simply \emph{FIFO}, from here on) would be a
weaker assumption (and therefore better) than synchrony.  FIFO in fact
is supported in practical, widely-used infrastructures such as TCP and
the Advanced Message Queuing Protocol, better known as AMQP
\shortcitep{AMQP}.  An even weaker assumption than ordered delivery
would be asynchronous communication without any kind of ordered
delivery.  Then, protocols in the language could be enacted directly
over the Internet and in highly resource-constrained settings such the
IoT.

Relying on any kind of ordered delivery guarantees from the
communication infrastructure naturally limits the kinds of
infrastructure upon which a protocol may be used to implement a
multiagent system.  More importantly, as we show below, ordering
guarantees are inadequate for ensuring the correct enactment of even
simple protocols.

Prima facie, there appears to be a simple motivation for using
 FIFO channels between agents, as illustrated by the
enactments of \pr~\ref{pr:Want+WillPay}.

\begin{example}[Want+WillPay]\label{pr:Want+WillPay}
  \rname{buyer} sends \mname{Want} (some item) and then
  \mname{WillPay} (some amount) to \rname{seller}.
\end{example}

\begin{figure}[!htb]
\centering
  \tikzstyle{role}=[thin,draw,align=center,font=\small\sffamily,rectangle,anchor=center,minimum height=5ex,minimum width=6ex,inner sep=1]

  \tikzstyle{every text node part/.style}=[align=center]

  \tikzset{every node text/.style={node
      contents=\transform{#1}}}
\newcommand{\transform}[1]{\ensuremath{\mathsf{#1}}}

  \tikzstyle{m_label_base}=[draw=none,midway,fill=none,sloped,align=center,font=\small\sffamily]
  \tikzstyle{m_label_up}=[m_label_base,above=-2pt]
  \tikzstyle{m_label_down}=[m_label_base,below=-2pt]

 \tikzstyle{a_label}=[draw=none,sloped,fill=white,align=center,font=\small\scshape]

% \tikzstyle{message}=[->, >=stealth']
  \tikzstyle{emptybox}=[draw=none,minimum height=3ex]

\begin{subfigure}[t]{0.48\columnwidth}
  \centering
\begin{tikzpicture}

\matrix () [row sep=10,column sep=60] {
  \node[emptybox] (a) {};
  &  \node[emptybox] (b) {};\\
  \node (a-zero) {};
  & \node (b-zero) {};\\
  \node (a-one) {};
  & \node (b-one) {};\\
  \node (a-two) {};
  & \node (b-two) {};\\
  \node (a-three) {};
  & \node (b-three) {};\\
  \node (a-four) {};
  & \node (b-four) {};\\
};

\node [role,draw=none,anchor=south] at (a) {Buyer};
\node [role,draw=none,anchor=south] at (b) {Seller};

\draw [dashed] (a.center)--(a-four.center);
\draw [dashed] (b.center)--(b-four.center);

\draw [message] (a-zero.center)--node [m_label_up] {Want} (b-one.center);

\draw [message] (a-two.center)--node [m_label_up] {WillPay} (b-three.center);

\end{tikzpicture}
\caption{\mname{Want} received before \mname{WillPay}.}
\label{fig:wbwp}
\end{subfigure}
\begin{subfigure}[t]{0.48\columnwidth}
  \centering
\begin{tikzpicture}

\matrix () [row sep=10,column sep=60] {
  \node[emptybox] (a) {};
  &  \node[emptybox] (b) {};\\
  \node (a-zero) {};
  & \node (b-zero) {};\\
  \node (a-one) {};
  & \node (b-one) {};\\
  \node (a-two) {};
  & \node (b-two) {};\\
  \node (a-three) {};
  & \node (b-three) {};\\
  \node (a-four) {};
  & \node (b-four) {};\\
};

\node [role,draw=none,anchor=south] at (a) {Buyer};
\node [role,draw=none,anchor=south] at (b) {Seller};

\draw [dashed] (a.center)--(a-four.center);
\draw [dashed] (b.center)--(b-four.center);

\draw [message] (a-zero.center)--node [m_label_up,pos=.25] {Want} (b-three.center);

\draw [message] (a-one.center)--node [m_label_down,pos=.35] {WillPay} (b-two.north);
\end{tikzpicture}
\caption{\mname{WillPay} received before \mname{Want}.}
\label{fig:wpbw}
\end{subfigure}
\caption{In the absence of ordering guarantees from the
  infrastructure, messages could become reordered.  If the
  infrastructure provided  FIFO delivery, then the enactment
  of Figure~\ref{fig:wpbw} would not be possible.}
\label{fig:Want+WillPay}
\end{figure}
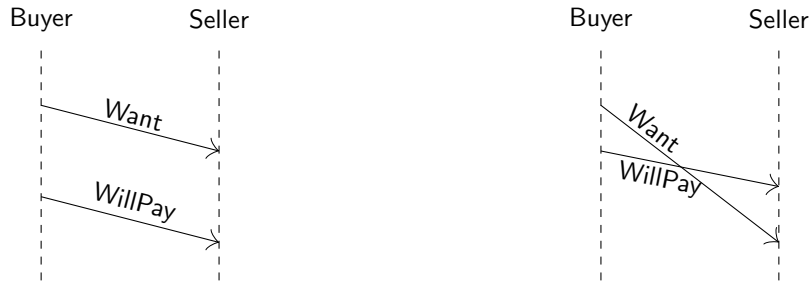

Figure~\ref{fig:Want+WillPay} shows two possible enactments of
\pr~\ref{pr:Want+WillPay}.  Notice that in Figure~\ref{fig:wpbw}, the
messages are reordered in transit so that \mname{WillPay} is received
by \rname{seller} before \mname{Want}.

Listing~\ref{Trace:Want+WillPay} gives the protocol in Trace-C and
Trace-F and its projections.

\begin{lstlisting}[label={Trace:Want+WillPay},caption={Want+WillPay in Trace-C and Trace-F along with projections.}]
  //Protocol Buyer $\ar{Want}$ Seller ; Buyer $\ar{WillPay}$ Seller

//Projections
Buyer: Seller!Want ; Seller!WillPay
Seller: Buyer?Want ; Buyer?WillPay
\end{lstlisting}

The protocol of Listing~\ref{Trace:Want+WillPay} cannot handle the
enactment of Figure~\ref{fig:wpbw} because the \rname{seller}'s
projection expects to receive \mname{Want} before \mname{WillPay}.
The enactment of Figure~\ref{fig:wpbw} though is ruled out if the
infrastructure guarantees FIFO delivery; only the enactment of
Figure~\ref{fig:wbwp} is possible under FIFO.  That is, in essence,
there are two mutually exclusive alternatives: (1) either declare the
enactment with the reordered messages to be valid and the protocol to
be unrealizable; or, (2) assume FIFO channels between agents and
declare the protocol to be realizable.  Scribble and Trace-C take the
latter alternative.  Trace-F also effectively takes the latter
alternative: the protocol is unrealizable under unordered asynchrony
but is realizable under FIFO (with the SR, SS, and RR
interpretations).

Listing~\ref{BSPL:Want+WillPay} gives a BSPL specification for
\pr~\ref{pr:Want+WillPay}.  The protocol ensures that \rname{buyer}
can send \mname{WillPay} with some binding for \paraname{ID} only
after \mname{Want}; however, it does not constrain when the messages
should be received by \rname{seller}.  Thus, it supports both
enactments of Figure~\ref{fig:Want+WillPay}.

\begin{lstlisting}[label={BSPL:Want+WillPay},caption={Want+WillPay in BSPL.}]
Want+WillPay {
 $\role$ Buyer, Seller
 $\param$ out ID, out item, out price

 Buyer $\mo$ Seller: Want[out ID, out item]
 Buyer $\mo$ Seller: WillPay[in ID, in item, out price]
}
\end{lstlisting}

Assuming FIFO to help ensure correctness seems innocuous at first
glance.  However, a FIFO communication infrastructure is infeasible in
important application settings.  For example, in the Internet of
Things (IoT), many of the devices lack a capability for anything
beyond packet-based communication and in particular lack the
capability for buffering.  Buffering is the common way to implement
FIFO; another implementation approach would be to drop a message whose
sequence number is not the next number to the one most recently
received---but that too is not practicable since it would waste
resources and exacerbate the latency of communication.  In settings
that demand fast interactions (e.g., for financial transactions), the
additional latency due to FIFO is an avoidable overhead.

Moreover, and crucially, the FIFO assumption faces a profound semantic
problem: FIFO turns out to be inadequate for correctness in settings
of more than two parties, as \pr~\ref{pr:indirect-payment}
demonstrates.

\begin{example}[Indirect payment]\label{pr:indirect-payment}
  In an indirect-payment purchase protocol, after receiving an
  \mname{Offer}, \rname{buyer} first sends \mname{Accept} to
  \rname{seller} and then sends \mname{Instruct} (a payment
  instruction) to \rname{bank}.  Upon receiving \mname{Instruct},
  \rname{bank} sends a funds \pname{Transfer} to \rname{seller}.
\end{example}

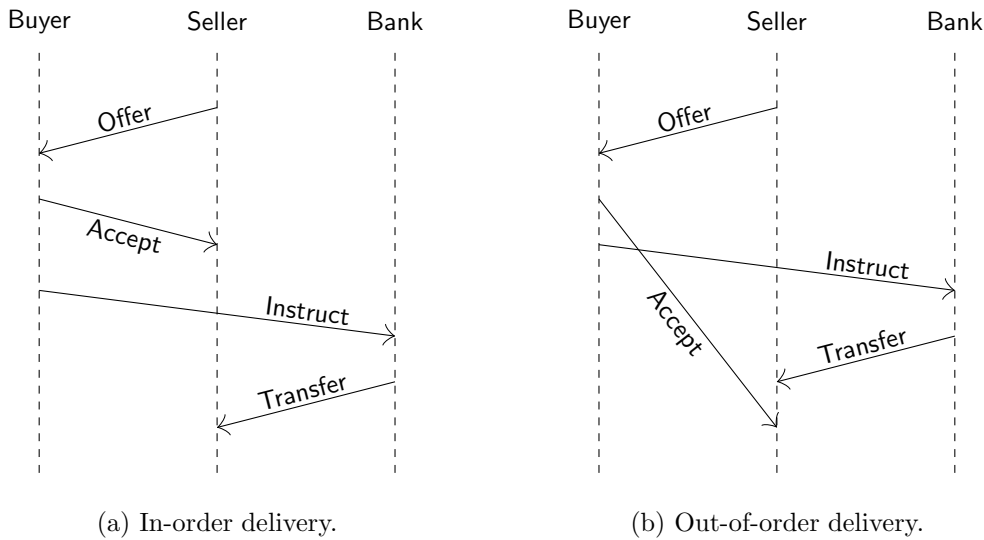
\begin{figure}[!htb]
\centering
  \tikzstyle{role}=[thin,draw,align=center,font=\small\sffamily,rectangle,anchor=center,minimum height=5ex,minimum width=6ex,inner sep=1]

  \tikzstyle{every text node part/.style}=[align=center]

  \tikzset{every node text/.style={node
      contents=\transform{#1}}}
\newcommand{\transform}[1]{\ensuremath{\mathsf{#1}}}

  \tikzstyle{m_label_base}=[draw=none,midway,fill=none,sloped,align=center,font=\small\sffamily]
  \tikzstyle{m_label_up}=[m_label_base,above=-2pt]
  \tikzstyle{m_label_down}=[m_label_base,below=-2pt]

 \tikzstyle{a_label}=[draw=none,sloped,fill=white,align=center,font=\small\scshape]

% \tikzstyle{message}=[->, >=stealth']
  \tikzstyle{emptybox}=[draw=none,minimum height=3ex]

\begin{subfigure}[t]{0.48\columnwidth}
  \centering
\begin{tikzpicture}

\matrix () [row sep=10,column sep=60] {
  \node[emptybox] (a) {};
  &  \node[emptybox] (b) {};
  &  \node[emptybox] (c) {};\\
  \node (a-zero) {};
  & \node (b-zero) {};
  & \node (c-zero) {}; \\
  \node (a-one) {};
  & \node (b-one) {};
  & \node (c-one) {}; \\
  \node (a-two) {};
  & \node (b-two) {};
  & \node (c-two) {}; \\
  \node (a-three) {};
  & \node (b-three) {};
  & \node (c-three) {}; \\
  \node (a-four) {};
  & \node (b-four) {};
  & \node (c-four) {}; \\
  \node (a-five) {};
  & \node (b-five) {};
  & \node (c-five) {}; \\
  \node (a-six) {};
  & \node (b-six) {};
  & \node (c-six) {}; \\
  \node (a-seven) {};
  & \node (b-seven) {};
  & \node (c-seven) {}; \\
  \node (a-eight) {};
  & \node (b-eight) {};
  & \node (c-eight) {}; \\
};

\node [role,draw=none,anchor=south] at (a) {Buyer};
\node [role,draw=none,anchor=south] at (b) {Seller};
\node [role,draw=none,anchor=south] at (c) {Bank};

\draw [dashed] (a.center)--(a-eight.center);
\draw [dashed] (b.center)--(b-eight.center);
\draw [dashed] (c.center)--(c-eight.center);

\draw [message] (b-zero.center)--node [m_label_up] {Offer} (a-one.center);

\draw [message] (a-two.center)--node [m_label_down] {Accept} (b-three.center);

\draw [message] (a-four.center)--node [m_label_up,pos=.75] {Instruct} (c-five.center);

\draw [message] (c-six.center)--node [m_label_up] {Transfer} (b-seven.center);
\end{tikzpicture}
\caption{In-order delivery.}
\label{fig:FIFO-in-order}
\end{subfigure}
\begin{subfigure}[t]{0.48\columnwidth}
  \centering
\begin{tikzpicture}
\matrix () [row sep=10,column sep=60] {
  \node[emptybox] (a) {};
  &  \node[emptybox] (b) {};
  &  \node[emptybox] (c) {};\\
  \node (a-zero) {};
  & \node (b-zero) {};
  & \node (c-zero) {}; \\
  \node (a-one) {};
  & \node (b-one) {};
  & \node (c-one) {}; \\
  \node (a-two) {};
  & \node (b-two) {};
  & \node (c-two) {}; \\
  \node (a-three) {};
  & \node (b-three) {};
  & \node (c-three) {}; \\
  \node (a-four) {};
  & \node (b-four) {};
  & \node (c-four) {}; \\
  \node (a-five) {};
  & \node (b-five) {};
  & \node (c-five) {}; \\
  \node (a-six) {};
  & \node (b-six) {};
  & \node (c-six) {}; \\
  \node (a-seven) {};
  & \node (b-seven) {};
  & \node (c-seven) {}; \\
  \node (a-eight) {};
  & \node (b-eight) {};
  & \node (c-eight) {}; \\
};

\node [role,draw=none,anchor=south] at (a) {Buyer};
\node [role,draw=none,anchor=south] at (b) {Seller};
\node [role,draw=none,anchor=south] at (c) {Bank};

\draw [dashed] (a.center)--(a-eight.center);
\draw [dashed] (b.center)--(b-eight.center);
\draw [dashed] (c.center)--(c-eight.center);

\draw [message] (b-zero.center)--node [m_label_up] {Offer} (a-one.center);

\draw [message] (a-two.center)--node [m_label_down] {Accept} (b-seven.center);

\draw [message] (a-three.center)--node [m_label_up,pos=.75] {Instruct} (c-four.center);

\draw [message] (c-five.center)--node [m_label_up] {Transfer} (b-six.center);
\end{tikzpicture}
\caption{Out-of-order delivery.}
\label{fig:FIFO-out-of-order}
\end{subfigure}
\caption{FIFO communication does not guarantee consistent ordering
  across a multiagent system with three or more agents.  Each of these
  enactments respects FIFO because at most one message occurs on each
  channel (between each pair of roles).  In
  Figure~\ref{fig:FIFO-out-of-order}, whereas for \rname{buyer},
  \mname{Accept} occurs before \mname{Instruct}, for \rname{seller},
  \mname{Accept} occurs after \mname{Transfer}, and therefore,
  logically, after \mname{Instruct}.}
\label{fig:FIFO-no-order}

\end{figure}

Figure~\ref{fig:FIFO-no-order} shows two enactments for
\pr~\ref{pr:indirect-payment}.  In Figure~\ref{fig:FIFO-in-order},
\rname{seller} receives \mname{Accept} before \mname{Transfer} whereas
in Figure~\ref{fig:FIFO-out-of-order}, \rname{seller} receives
\mname{Accept} after \mname{Transfer}.  Both enactments satisfy FIFO
since at most one message is being sent on any channel.  The
enactments illustrate that even with FIFO ordering, asynchrony makes
ordering indeterminate for protocols involving more than two agents.

Listing~\ref{Trace:no-order} is an attempt to capture
\pr~\ref{pr:indirect-payment} in Trace-C.  Following the reasoning for
Trace-C \citep[p.~16]{Castagna+12:sessions}, this protocol is
unrealizable.  Specifically, the projection for \rname{seller} expects
to receive \mname{Accept} before \mname{Transfer} and therefore does
not support the enactment in Figure~\ref{fig:FIFO-out-of-order}, which
may arise despite using FIFO channels.  In summary, by ruling out the
protocol, Trace-C rules out realistic message orders that are simply
the result of asynchrony.  Listing~\ref{Trace:no-order} additionally
serves as the specification of the protocol in Trace-F.  Under FIFO,
no matter what sequence interpretation is chosen, the protocol is
unrealizable.

\begin{lstlisting}[label={Trace:no-order},caption={Indirect payment (\pr~\ref{pr:indirect-payment}) protocol and its projections in Trace-C and Trace-F.}]
//Indirect payment
Seller $\ar{Offer}$ Buyer; Buyer $\ar{Accept}$ Seller;
Buyer $\ar{Instruct}$ Bank; Bank $\ar{Transfer}$ Seller

//Projections
Buyer: Seller?Offer; Seller!Accept; Bank!Instruct
Seller: Buyer!Offer; Buyer?Accept; Bank?Transfer
Bank: Buyer?Instruct; Seller!Transfer
\end{lstlisting}

Listing~\ref{Scribble:no-order} gives a Scribble protocol to capture
\pr~\ref{pr:indirect-payment}.  The Scribble projections are analogous
to the Trace-C and Trace-F projections in
Listing~\ref{Trace:no-order}.  In particular, \rname{seller} cannot
receive \mname{Transfer} before \mname{Accept}; it \emph{blocks} on
the reception of \mname{Accept} on the channel from \rname{buyer} even
when \mname{Transfer} may have arrived earlier on the channel from
\rname{bank}.  The listing shows \rname{seller}'s projection (other
agents' projections are elided).  \review{B:reorders}Effectively, the
projection enforces a reception order for the two messages that may be
different from their arrival order.

\begin{lstlisting}[label={Scribble:no-order},caption={Indirect payment in Scribble.}]
global protocol IndirectPayment(role Buyer, role Seller, role Bank) {
  Offer() from Seller to Buyer;
  Accept() from Buyer to Seller;
  Instruct() from Buyer to Bank;
  Transfer() from Bank to Seller;
}

local protocol IndirectPayment_Seller(role Buyer, role Seller, role Bank) {
  Offer() to Buyer;
  Accept() from Buyer;
  Transfer() from Bank;
}
\end{lstlisting}

The BSPL specification in Listing~\ref{BSPL:no-order} specifies a
protocol that supports both of the enactments shown in
Figure~\ref{fig:FIFO-no-order}.  The reason is that, in BSPL, an agent
may receive a message whenever the communication infrastructure
delivers a message to the agent.  Information causality and integrity
in BSPL constrain the emission of messages by an agent; message
reception is unconstrained.

\begin{lstlisting}[label={BSPL:no-order},caption={Indirect payment protocol in BSPL.}]
Indirect Payment {
  $\role$ Buyer, Seller, Bank
  $\param$ out ID key, out item, out price, out decision, out instruction, out OK

  Seller $\mo$ Buyer: Offer[out ID, out item, out price]
  Buyer $\mo$ Seller: Accept[in ID, in item, in price, out decision]
  Buyer $\mo$ Bank: Instruct[in ID, in price, in decision, out instruction]
  Bank $\mo$ Seller: Transfer[in ID, in price, in instruction, out OK]
}
\end{lstlisting}

We omit HAPN from this discussion since it does not support
asynchrony.

\section{Mapping Back to Multiagent Systems}
\label{sec:mapping}

We now map the findings from our analysis of the protocol languages to
multiagent systems.

\subsection{Architecture}
We identify the architectural assumptions that underlie the protocol
languages discussed above.

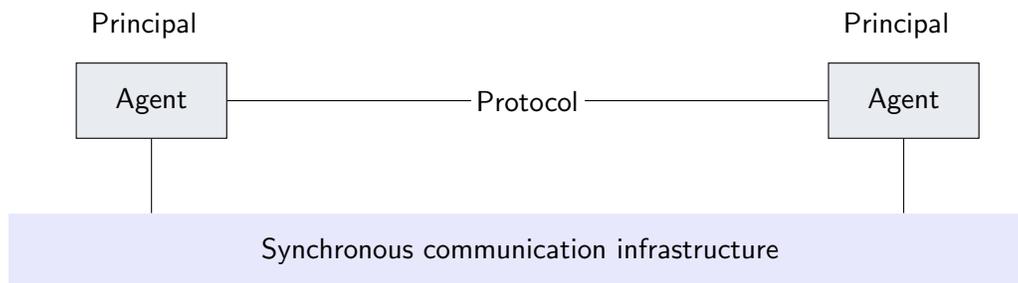
\begin{figure}[ht]
  \centering
  \begin{tikzpicture}[>=stealth]

\tikzstyle{box}=[draw=none,rounded corners,align=center,font=\sffamily,fill=blue!20!gray!80,rectangle,anchor=center,minimum height=4ex,minimum width=12ex,inner sep=2] %,text width=13ex

\tikzstyle{pbox}=[draw=none,sharp corners,align=center,font=\sffamily,fill=blue!20!gray!40,rectangle,anchor=center,minimum height=6ex,minimum width=20ex,inner sep=2]

\tikzstyle{cbox}=[sharp corners,align=center,font=\sffamily,fill=blue!80!gray!10,rectangle,anchor=center,minimum height=6ex,minimum width=82ex,inner sep=2]

\tikzstyle{ebox}=[draw=none,sharp corners,align=center,font=\sffamily,fill=blue!20!gray!40,rectangle,anchor=center,minimum height=3.5ex,minimum width=12ex,inner sep=2]

\tikzstyle{dmbox}=[draw=none,sharp corners,align=center,font=\sffamily,fill=blue!20!gray!40,rectangle,anchor=center,minimum height=15ex,minimum width=6ex,inner sep=2]

\tikzstyle{lane}=[draw=none]

\tikzstyle{bar}=[draw=none,fill=black,minimum width=8ex]

\tikzstyle{edge_label}=[draw=none, pos=0, fill=white,inner sep=2,font=\sffamily,align=center]
\tikzstyle{west_label}=[edge_label] %,anchor=west
\tikzstyle{east_label}=[edge_label] %,anchor=east

\tikzstyle{edge_label_org}=[draw=none, pos=0, fill=white,inner
sep=2,font=\sffamily] % ,align=center,y

\tikzstyle{every text node part/.style}=[align=center]

\tikzset{
  big arrow/.style={
    decoration={markings,mark=at position 1 with
%% {\arrow[scale=2,#1]{>}}
{\arrow[line width=2.5,#1]{>}}
},
    postaction={decorate},
    shorten >=0.4pt},
  big arrow/.default=blue}

\tikzset{
  big arrow LR/.style={
    decoration={markings,
mark=at position 6pt with {\arrow[line width=2.5,#1]{<}};,
mark=at position 1 with {\arrow[line width=2.5,#1]{>}};},
    postaction={decorate},
    shorten <=1pt,
    shorten >=0.4pt},
  big arrow LR/.default=blue}

\tikzset{darkarrow/.style={big arrow,colorEntityBack,thick}, big arrow/.default=colorEntityBack}

\tikzset{bluearrowLR/.style={big arrow LR,blue, thick}, big
  arrow LR/.default=blue}

\node[pbox,fill=listbackgroundcolorlight] (la) at (-1,0) [draw,minimum width=2cm,minimum height=1cm] {Agent};
\node[pbox,fill=listbackgroundcolorlight] (ra) at (9,0) [draw,minimum width=2cm,minimum height=1cm] {Agent};
\node[box,fill=none] (lp) at ($(la.north)+(-0.1,.5)$) {Principal};
\node[box,fill=none] (rp) at ($(ra.north)+(-0.1,.5)$) {Principal};

\draw  (la)--node [edge_label_org,midway, align=center] (protocol) {Protocol} (ra);

\node[cbox](cic) at  (3.9,-2) {Synchronous communication infrastructure};
\draw (la)--($(cic.north)+(-4.9,0)$);
\draw (ra)--($(cic.north)+(5.1,0)$);
\end{tikzpicture}

\caption{Architecture induced by HAPN.  Communication between agents
  is synchronous.}
\label{fig:HAPN-MAS}
\end{figure}

\review{B:H-two} Figure~\ref{fig:HAPN-MAS} depicts the architecture
underlying HAPN.  HAPN does not give a method for deriving projections
from protocols; assuming, however, a suitable method for projecting
and enacting HAPN protocols, the major difference from the
architecture in Figure~\ref{fig:minimal-MAS} is that HAPN, as
currently formalized, requires synchronous communication, a violation
of Constraints~\ref{req:nonblocking-emission}
and~\ref{req:anytime-reception}.

\begin{figure}[ht]
  \centering
  \begin{tikzpicture}[>=stealth]

\tikzstyle{box}=[draw=none,rounded corners,align=center,font=\sffamily,fill=blue!20!gray!80,rectangle,anchor=center,minimum height=4ex,minimum width=12ex,inner sep=2] %,text width=13ex

\tikzstyle{pbox}=[draw=none,sharp corners,align=center,font=\sffamily,fill=blue!20!gray!40,rectangle,anchor=center,minimum height=6ex,minimum width=20ex,inner sep=2]

\tikzstyle{cbox}=[sharp corners,align=center,font=\sffamily,fill=blue!80!gray!10,rectangle,anchor=center,minimum height=6ex,minimum width=82ex,inner sep=2]

\tikzstyle{ebox}=[draw=none,sharp corners,align=center,font=\sffamily,fill=blue!20!gray!40,rectangle,anchor=center,minimum height=3.5ex,minimum width=12ex,inner sep=2]

\tikzstyle{dmbox}=[draw=none,sharp corners,align=center,font=\sffamily,fill=blue!20!gray!40,rectangle,anchor=center,minimum height=15ex,minimum width=6ex,inner sep=2]

\tikzstyle{lane}=[draw=none]

\tikzstyle{bar}=[draw=none,fill=black,minimum width=8ex]

\tikzstyle{edge_label}=[draw=none, pos=0, fill=white,inner sep=2,font=\sffamily,align=center]
\tikzstyle{west_label}=[edge_label] %,anchor=west
\tikzstyle{east_label}=[edge_label] %,anchor=east

\tikzstyle{edge_label_org}=[draw=none, pos=0, fill=white,inner
sep=2,font=\sffamily] % ,align=center,y

\tikzstyle{every text node part/.style}=[align=center]

\tikzset{
  big arrow/.style={
    decoration={markings,mark=at position 1 with
%% {\arrow[scale=2,#1]{>}}
{\arrow[line width=2.5,#1]{>}}
},
    postaction={decorate},
    shorten >=0.4pt},
  big arrow/.default=blue}

\tikzset{
  big arrow LR/.style={
    decoration={markings,
mark=at position 6pt with {\arrow[line width=2.5,#1]{<}};,
mark=at position 1 with {\arrow[line width=2.5,#1]{>}};},
    postaction={decorate},
    shorten <=1pt,
    shorten >=0.4pt},
  big arrow LR/.default=blue}

\tikzset{darkarrow/.style={big arrow,colorEntityBack,thick}, big arrow/.default=colorEntityBack}

\tikzset{bluearrowLR/.style={big arrow LR,blue, thick}, big
  arrow LR/.default=blue}

\node[pbox,fill=listbackgroundcolorlight] (la) at (-1,0) [minimum width=2cm,minimum height=1cm] {Agent};
\node[pbox,fill=listbackgroundcolorlight] (ra) at (9,0) [minimum width=2cm,minimum height=1cm] {Agent};

\draw  (la)--node [edge_label_org,midway, align=center] (protocol) {Protocol} (ra);

\node[cbox](cic) at  (3.9,-2) {FIFO-based asynchronous communication infrastructure};

\draw (la) to ($(cic.north)+(-4.9,0)$);
\draw (ra) to ($(cic.north)+(5.1,0)$);
\end{tikzpicture}

\caption{Architecture induced by Trace-C.  Communication is
  asynchronous via FIFO channels.  The architecture for Trace-F under
  FIFO-asynchrony is identical.}
\label{fig:Trace-MAS}
\end{figure}
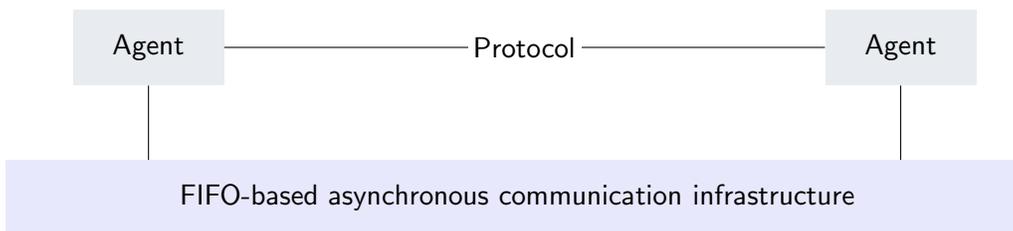

Figure~\ref{fig:Trace-MAS} shows the architecture induced by Trace-C.
Communication between agents is via a FIFO-based asynchronous
infrastructure.  Recall that Trace-F has a pluggable communication
infrastructure.  Figure~\ref{fig:Trace-MAS} also depicts the
architecture induced by Trace-F when FIFO is assumed.

\review{B:infra}FIFO (delivery) though is a stronger assumption
than Assumption~\ref{req:infrastructure}, which requires only
noncreativity from the infrastructure.  Without FIFO, Trace-C and
Trace-F would violate reception correctness
(Constraint~\ref{req:reception-correctness}), which states that any
reception of a message that is correctly sent is correct.  The Trace-C
and Trace-F \pname{Want+WillPay} protocol
(Listing~\ref{Trace:Want+WillPay}) illustrates the violation.  Even
though \rname{buyer} sends \mname{Want} before \mname{WillPay}, as
required by the protocol, \rname{seller} receiving \mname{WillPay}
before \mname{Want} is an incorrect enactment.

In fact, even with FIFO, both Trace-C and Trace-F violate
\emph{reception correctness}
(Constraint~\ref{req:reception-correctness}).  This is illustrated by
the Trace-C and Trace-F \pname{Indirect Payment} protocol
(Listing~\ref{Trace:no-order}).  Asynchrony means that
\mname{Transfer} may be received by \rname{seller} before
\mname{Accept}; however, that order of reception is incorrect
according to the protocol.

\begin{figure}[ht]
  \centering

  \begin{tikzpicture}[>=stealth]

\tikzstyle{box}=[draw=none,rounded corners,align=center,font=\sffamily,fill=blue!20!gray!80,rectangle,anchor=center,minimum height=4ex,minimum width=12ex,inner sep=2] %,text width=13ex

\tikzstyle{pbox}=[draw=none,sharp corners,align=center,font=\sffamily,fill=blue!20!gray!40,rectangle,anchor=center,minimum height=6ex,minimum width=20ex,inner sep=2]

\tikzstyle{cbox}=[sharp corners,align=center,font=\sffamily,fill=blue!80!gray!10,rectangle,anchor=center,minimum height=6ex,minimum width=82ex,inner sep=2]

\tikzstyle{ebox}=[draw=none,sharp corners,align=center,font=\sffamily,fill=blue!20!gray!40,rectangle,anchor=center,minimum height=3.5ex,minimum width=12ex,inner sep=2]

\tikzstyle{dmbox}=[draw=none,sharp corners,align=center,font=\sffamily,fill=blue!20!gray!40,rectangle,anchor=center,minimum height=15ex,minimum width=6ex,inner sep=2]

\tikzstyle{lane}=[draw=none]

\tikzstyle{bar}=[draw=none,fill=black,minimum width=8ex]

\tikzstyle{edge_label}=[draw=none, pos=0, fill=white,inner sep=2,font=\sffamily,align=center]
\tikzstyle{west_label}=[edge_label] %,anchor=west
\tikzstyle{east_label}=[edge_label] %,anchor=east

\tikzstyle{edge_label_org}=[draw=none, pos=0, fill=white,inner
sep=2,font=\sffamily] % ,align=center,y

\tikzstyle{every text node part/.style}=[align=center]

\tikzset{
  big arrow/.style={
    decoration={markings,mark=at position 1 with
%% {\arrow[scale=2,#1]{>}}
{\arrow[line width=2.5,#1]{>}}
},
    postaction={decorate},
    shorten >=0.4pt},
  big arrow/.default=blue}

\tikzset{
  big arrow LR/.style={
    decoration={markings,
mark=at position 6pt with {\arrow[line width=2.5,#1]{<}};,
mark=at position 1 with {\arrow[line width=2.5,#1]{>}};},
    postaction={decorate},
    shorten <=1pt,
    shorten >=0.4pt},
  big arrow LR/.default=blue}

\tikzset{darkarrow/.style={big arrow,colorEntityBack,thick}, big arrow/.default=colorEntityBack}

\tikzset{bluearrowLR/.style={big arrow LR,blue, thick}, big
  arrow LR/.default=blue}

\node[pbox,fill=listbackgroundcolorlight] (la) at (-1,0) [minimum width=2cm,minimum height=1cm] {Agent};
\node[pbox,fill=listbackgroundcolorlight] (ra) at (9,0) [minimum width=2cm,minimum height=1cm] {Agent};

\node[pbox,fill=listbackgroundcolorlight] (l-real) at (-1,-2) [minimum width=2cm,minimum height=1cm] {Channel\\selector};
\node[pbox,fill=listbackgroundcolorlight] (r-real) at (9,-2) [minimum width=2cm,minimum height=1cm] {Channel\\selector};

\draw  (la)--node [edge_label_org,midway, align=center] (protocol) {Protocol} (ra);
\draw  (l-real)--node [edge_label_org,midway, align=center] (protocol) {Protocol} (r-real);

\node[cbox](cic) at  (3.9,-4) {FIFO-based asynchronous communication infrastructure};
\draw (la)--(l-real);
\draw (ra)--(r-real);
\draw (l-real) to ($(cic.north)+(-4.9,0)$);
\draw (r-real) to ($(cic.north)+(5.1,0)$);
\end{tikzpicture}
\caption{Architecture induced by Scribble.  Communication between
  agents is via FIFO channels.  At any time, the channel selector
  hides all channels from the agent except the one on which the
  message expected by an agent at that time will arrive.}
\label{fig:Scribble-MAS}
\end{figure}
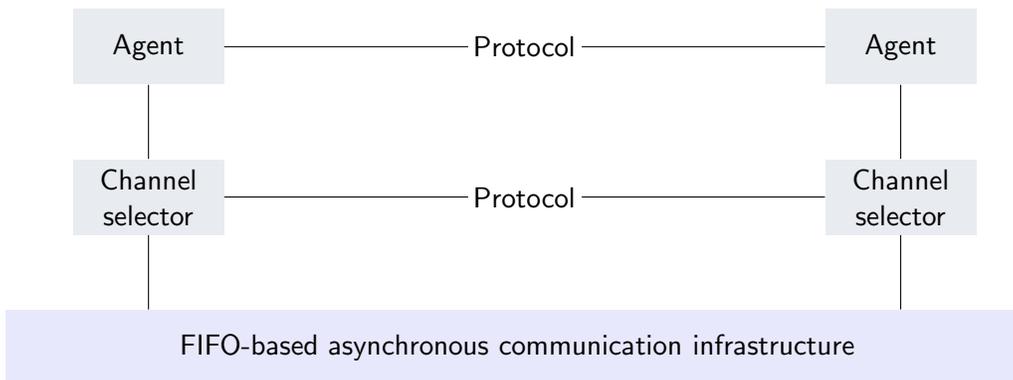

Scribble induces the architecture in Figure~\ref{fig:Scribble-MAS}.
As with the architecture for Trace-C in Figure~\ref{fig:Trace-MAS}, it
requires FIFO.  If FIFO were to be dropped, Scribble, like Trace-C and
Trace-F, would violate reception correctness
(Constraint~\ref{req:reception-correctness}).

Scribble treats the reception of a message as a logically blocking
operation on the channel on which the message is expected: an agent
will not receive a message on any other channel until it receives the
message it is blocked on.  The notion of blocking reception is
instrumental to Scribble's ability to handle the arrival of
\mname{Transfer} before the arrival of \mname{Accept} in the indirect
payment protocol.  Specifically, \rname{seller} blocks on the channel
from \rname{buyer}, on which \mname{Accept} is expected; in the
meantime, if \mname{Transfer} arrives on the channel from the banker,
it is ignored and thereby not considered received.

In essence, Scribble reorders message arrivals to suit the agent.
Such reordering is incompatible with asynchrony; specifically, it is a
violation of anytime reception
(Constraint~\ref{req:anytime-reception}).  Scribble attempts to
disguise the violation by treating the reception of a message as the
result of an agent's decision to receive the message, distinct from
the event of the arrival of the message.  Architecturally, the
reordering across channels is captured in the \emph{channel selector}
component in Figure~\ref{fig:Scribble-MAS} that mediates between the
FIFO infrastructure and an agent and hides all channels from the agent
except the one on which it is expecting to receive a message.

If the idea of reordering messages received on different channels were
dropped, then even with FIFO, Scribble would violate reception
correctness (Constraint~\ref{req:reception-correctness}) (just as
Trace-C and Trace-F do).

By contrast, Figure~\ref{fig:minimal-MAS} captures the architecture
induced by BSPL.  A multiagent system based on BSPL satisfies all the
constraints given in Section~\ref{sec:intro}.  In particular, BSPL
works with asynchronous communication but does not require FIFO.
Given any safe BSPL protocol, any message sent according to the
protocol is also received correctly regardless of when that reception
occurs relative to other receptions.

For completeness, we comment on some recent programming and architectural
frameworks for MAS.
% JADE \citep{bellifemine:jade:2007}, one of the earliest programming
% frameworks for multiagent systems, supports the FIPA Interaction
% Protocols \citep{FIPA-IP}.  However, as mentioned earlier, the
% protocols are specified in AUML an informal notation, which limits
In JaCaMo \citep{boissier:jacamo:2013}, agents coordinate
their computations by shared \emph{artifacts}---components that
provide ``functionalities and services'' (p.~750) for agents.  JaCaMo
can be used to realize a MAS that satisfies the canonical
architectural style of Figure~\ref{fig:minimal-MAS}---or any of the
others for that matter.  Specifically, messaging between agents could
be realized using artifacts.  The infrastructure can avoid
centralization by, for example, using a dedicated artifact for each
channel and deploying the artifacts across the system.  The sender of
a message would send it to the channel artifact and the receiver would
pick it from that artifact.  Indeed \citet{baldoni:artifacts:2019}
implement commitment-based coordination between agent by representing
commitments in a shared JaCaMo artifact.

Jason \citep{Bordini+07}, the language in which agents are programmed
in JaCaMo, supports specifying an agent's reasoning about incoming and
outgoing messages.  However, Jason's programming model does not
include protocols of the sort motivated here.

ReST \citep{Vinoski-08} is an architectural style for Web applications
that has been advocated as a basis for engineering MAS
\citep{Ciortea+18:manufacturing}.  An important ReST constraint is
that an application's state is fully captured in the representation of
the relevant Web resources (as identified by URIs) on the server.  The
architectural style that BSPL adopts is analogous to ReST but in a
more general peer-peer setting \citep{ICWS-LoST-11}.

\subsection{Principles for MAS}
\label{sec:principles}

We present some broad principles that are relevant for MAS but are
violated by several of the evaluated protocol languages.

\begin{principle}[No unitary
  perspective]\label{pr:no-unitary}
  A protocol should avoid specifying computations (enactments) from a
  unitary perspective.
\review{C:principle}
\end{principle}

Principle~\ref{pr:no-unitary} follows from the fact that there is no
valid unitary perspective in a MAS \citep{Hewitt+73,Gasser-91}; the
only perspectives that count are those of the agents in the MAS.
Protocols that specify computations from a unitary perspective are
either (1) unrealizable by agents acting based solely on their local
knowledge or (2) unduly restrict concurrency.

Scribble, Trace-C, Trace-F, and HAPN violate
Principle~\ref{pr:no-unitary}.
\review{C:unit}
In each of those languages, a protocol's computations are given from a
unitary perspective.   HAPN assumes synchrony \citep[p.~61]{Winikoff+18:HAPN} and does not
provide a method to generate endpoint projections.  Some HAPN
protocols may not be realizable without the assumption of synchrony.
\review{B:H-one}
A computation of a Scribble, Trace-C, or Trace-F protocol is a single
sequence of messaging events.  Scribble, Trace-C, and Trace-F support
asynchrony and each describes how to derive the local perspective,
that is, the projection, for each agent.  In the Scribble, Trace-C,
and Trace-F approaches, extracting the projections relies on a custom
(and usually complicated) theory of causality.
\review{B:failure}
Despite all the machinery that goes into
specifying the semantics of protocols, projections, and realizability
in these approaches, as we saw above, they fail to model important
aspects of the simple and realistic uses cases described above.

The Scribble, Trace-C, and Trace-F protocols for \mname{Flexible
  Purchase} (in Section~\ref{sec:concurrency}) highlight the pitfalls
of the unitary perspective.  From the unitary perspective, as
Listing~\ref{Trace-C:concurrency-transform} illustrates, it is a clear
choice between either \mname{Payment} first or \mname{Shipment} first.
However, when \rname{seller} and \rname{buyer} exercise their
respective choices locally, that is, on the basis of their
projections, a deadlock may obtain.  This leads to the protocols being
ruled out as unrealizable.

Protocols in BSPL satisfies Principle~\ref{pr:no-unitary}: The
computations of a BSPL protocol are given directly in terms of agent
perspectives.

Any protocol is a ``global'' specification for a MAS in the sense that
it specifies the (public) computations of the entire MAS.  And every
protocol conceptually yields a ``local'' specification, or projection,
for every role in the protocol.
\review{C:unitary}
Such a global-local distinction is
perfectly reasonable and not in conflict with
Principle~\ref{pr:no-unitary}.
\review{B:unitary}
Indeed, the global-local distinction
applies to BSPL as well: A BSPL protocol is a global specification in
the foregoing sense and a projection for a role in the protocol is the
set of all message specifications referenced by the protocol where the
role is either sender or receiver.  What Principle~\ref{pr:no-unitary}
rules out is giving the computations of a global protocol from a
unitary perspective.

In the discussion of the following principles, Trace-F stands for
Trace-F under FIFO asynchrony.  We can ignore Trace-F under other
communication models for two reasons.  First, under one of those
models, namely, unordered asynchrony, Trace-F is too weak a
language---it is unable to capture a protocol as simple as
Want+WillPay (\pr~\ref{pr:Want+WillPay}).  Second, all the
remaining communication models are strictly stronger than FIFO and the
conclusions we draw for Trace-F under FIFO asynchrony are valid for
them as well.

\begin{principle}[Noninterference]\label{pr:noninterference}
  A protocol must not prevent legitimate agent reasoning.
\end{principle}

An agent may want to process a message as soon as the message has
arrived.  The agent's motivations for doing so need not concern us
since the agents are autonomous and may adopt any preferences
arbitrarily in light of their autonomy.  However, Scribble, Trace-C,
and Trace-F, in requiring messages to be received in a certain other
relative to other messages, rule out such agents and, therefore,
violate Principle~\ref{pr:noninterference}.

For example, in \pr~\ref{pr:Want+WillPay}, \rname{seller} may want to
process the \mname{WillPay} message even if \mname{Want} hasn't yet
been received.  In \pr~\ref{pr:indirect-payment}, \rname{seller} may
want to process the \mname{Transfer} message even if \mname{Accept}
hasn't been received.  The Scribble, Trace-C, and Trace-F protocols
rule out this possibility.  BSPL, by contrast, satisfies
Principle~\ref{pr:noninterference}: the only constraints it imposes on
agents have to do with information causality and information
integrity.  Indeed, the BSPL protocols for \pr~\ref{pr:Want+WillPay}
and \pr~\ref{pr:indirect-payment} allow \rname{seller} to process
\mname{WillPay} before \mname{Want} is received and \mname{Transfer}
before \mname{Accept} is received, respectively.

\begin{principle}[End-to-end principle for protocols]\label{pr:E2E-protocols}
  Correct protocol enactment must not rely on message ordering
  guarantees from the communication infrastructure since the
  appropriate constraints are to be implemented and checked in
  agents.
\end{principle}

Scribble, Trace-C, and Trace-F violate Principle~\ref{pr:E2E-protocols} because
they require a FIFO communication infrastructure.  BSPL
satisfies Principle~\ref{pr:E2E-protocols} because it requires no ordering
guarantees from the infrastructure.

Principle~\ref{pr:E2E-protocols} derives from the end-to-end argument
for system design \citep{Saltzer+84}, which originated in the early
days of computer networking in the 1960s.  In its networking form, the
end-to-end principle states that functionality that makes sense at a
higher (``application'') layer should not be replicated at a lower
(``infrastructure'') layer.  The motivation is that if some function
of a networked application can be fully and correctly implemented only
at the application's end points standing above a communication
infrastructure, then supporting that function partially in the
infrastructure (1) is not adequate (still requires the application to
work to achieve that function); (2) imposes a cost on all internal
nodes; and (3) restricts the functioning of application end points
that are not concerned with that function.  We can think of FIFO as a
``function'' in this case---notably, \citeauthor{Saltzer+84}
specifically discuss the disadvantages of adopting a FIFO delivery
infrastructure; such an infrastructure is also not assumed in the
actor model \citep{Agha86}.

A protocol can be fully and correctly implemented only by the agents
who instantiate it.  Relying on the infrastructure for correctness is
of little benefit.  For explanation, let's consider emissions and
receptions, the two kinds of observations that a protocol may
constrain.  An agent must ensure the correctness of its emissions
because emissions are driven by an agent's internal decision-making.
For the correctness of receptions, an agent may rely on ordering
guarantees from the infrastructure.  However, such guarantees may be
insufficient for correctness.  For example, as we saw in the modeling
of indirect payment (\pr~\ref{pr:indirect-payment}), Scribble,
Trace-C, and Trace-F's reliance on FIFO turns out be insufficient for
correctness.

In addition to being insufficient, infrastructure ordering guarantees
may be \emph{excessive} since they constrain even messages that are
unrelated in terms of meaning but merely contingently happen to occur
together.  Listing~\ref{BSPL:unrelated} is illustrative of how
infrastructure ordering may be excessive.  The listing specifies two
protocols \pname{Just-Want} and \pname{Hello-World}.  Each protocol
has a single message from \rname{buyer} to \rname{seller}; the two
messages do not have any parameters in common, signifying that there
is no causal dependency between them.  Even so, a FIFO infrastructure
will necessarily deliver the messages to \rname{seller} in the order
in which they are sent by \rname{buyer}.

\begin{lstlisting}[label={BSPL:unrelated},caption={BSPL protocols that
    illustrate that ordering guarantees provided by the infrastructure
    may be excessive from the point of view of coordination.
    Although the messages in the two protocols are unrelated, if the
    infrastructure were FIFO, the messages would necessarily be delivered in
    the order sent.}]
Just-Want {
 $\role$ Buyer, Seller
 $\param$ out ID, out item

 Buyer $\mo$ Seller: Want[out ID, out item]
}

Hello-World  {
 $\role$ Buyer, Seller
 $\param$ out gID, out utterance

 Buyer $\mo$ Seller: Greeting[out gID, out utterance]
}
\end{lstlisting}

Finally, consider that ordering guarantees from the infrastructure come
with a heavy price: increased complexity and overhead in the
architecture, restrictions on the settings in which a multiagent
application may be deployed, and interference with higher-level agent
reasoning (as we discussed in the context of
Principle~\ref{pr:noninterference}).

\section{Discussion}
\label{sec:discussion}

Our contribution in this paper is an evaluation of select modern,
formal, and prominent languages for specifying protocols.  Our
evaluation criteria have to do with representations and operational
assumptions.  Our evaluation is concrete and comparative, driven by
the specification of protocols in the selected approaches, followed by
an analysis of the specifications.  The Scribble, Trace-F, and BSPL
protocols have been verified in their respective tooling.  We
understand verification tools for Trace-C and HAPN are not available.

Table~\ref{tab:comparisons} summarizes our findings.  For reasons
given above, Trace-F stands for Trace-F under FIFO-asynchrony.  Our
evaluation shows that BSPL is able to model the use cases considered
in all their richness despite---or because of---weaker guarantees from
the communication infrastructure.

\newcommand{\hdr}[1]{\multicolumn{5}{l}{\textcolor{blue!60!black}{\fsf{#1}}}}

\begin{table}[htb]
  \caption{Summary of evaluation.  The table indicates for each
    language, whether it fully satisfies (Yes), partially satisfies
    (Partial), or does not satisfy (No) each criterion.}
\label{tab:comparisons}
\centering
\begin{tabular}{l l c c c c c}\toprule
% \begin{tabular}{@{~}l@{~~} c@{~~} c@{~~} c@{~~} c@{~~} c@{~}}\toprule
& \fsf{Criterion} & \fsf{Scribble}& \fsf{Trace-C} & \fsf{Trace-F} & \fsf{HAPN} & \fsf{BSPL} \\
\midrule %
\hdr{Information} \\
  & Instances & Partial & Partial & Partial & Partial & Yes \\
  & Integrity & No & No & Partial & Partial & Yes \\
  & Social meaning & Partial & Partial & Partial & Partial & Yes \\
\hdr{Flexibility}\\ %
  & Concurrency & No & No & No & No & Yes \\
  & Extensibility & No & No &  No & No & Yes \\
\hdr{Operational environment} \\
  & Asynchrony & Yes & Yes & Yes & No & Yes \\
  & Unordering & No & No & No & No & Yes
  \\ \bottomrule
\end{tabular}
\end{table}

We discussed how Scribble, Trace-C, Trace-F, and HAPN violate the
canonical MAS architectural style presented in
Section~\ref{sec:intro}.  HAPN requires synchrony, which makes it
impractical for decentralized settings.  Scribble, Trace-C, and
Trace-F reorder messages to fit an agent's perspective.  BSPL works
with unordered asynchrony; the MAS architecture induced by BSPL is
compatible with the canonical architectural style.

\review{B:immutable}
Information (as captured by parameter bindings) in BSPL are immutable.
For MAS where agents communicate asynchronously, the immutability of
information simplifies ensuring correctness even though the
participants make observations in different orders---as expressed by
\citet{chandy:snapshots:1985}.

The immutability of information, however, calls for a different way of
specifying interactions.  Specifically, to achieve the effect of
updates, a BSPL protocol must incorporate a composite key such that a
part of the key reflects the version ID, thereby preventing the
versions from interfering with each other.  To capture finality, the
protocol must incorporate a message that conflicts with (and thus
prevents) subsequent versions.

In this way, there is a tradeoff between (1) adopting a new way of
modeling interaction to accommodate realistic communication
assumptions (asynchrony), and (2) making unrealistic assumptions
(synchrony) to support a familiar programming style (mutability).
Immutability in BSPL contrasts most clearly with HAPN, which
explicitly supports both bind and unbind operations on parameters and
therefore supports rebinding a parameter.

We set out to evaluate protocol languages from the standpoint of
building decentralized  multiagent systems.
\review{D:results}
An informally but widely held attitude in the multiagent systems
research community (and in related external subcommunities) is that
our approaches are adequate for tackling autonomy and flexibility as
needed in decentralized settings.
\review{E:results}
Accordingly, researchers have taken
support for decentralization as a done deal and gone on to tackle
challenges such as tool support and ease of use---which are no doubt
crucial challenges.  However, as we have shown in this paper, the more
fundamental challenges of decentralization have largely not been
overcome by languages of the established paradigms.  One language from
a new, information-oriented paradigm does tackle those challenges.
Whether that or another, as yet unknown, paradigm prevails remains to
be seen.  But what we can conclude from this exercise is that a new
way of thinking is needed to properly accommodate decentralization.

\subsection{Other Criteria for Evaluating Protocol Languages}
\label{Sec:other-criteria}

The criteria by which we study protocol languages
in this paper are not exhaustive.
\review{B:criteria}
\citet{Winikoff+18:HAPN} compare
several protocol languages, including HAPN and BSPL, for orthogonal
criteria such as precision, simplicity, graphical representation, and
so on.
\review{C:multiple}
They note that whereas HAPN supports
multiple agents playing a role, BSPL doesn't (Splee
\citep{AAMAS-17:Splee}, an extension of BSPL, though supports this
feature).  \citeauthor{Winikoff+18:HAPN} also report on their personal
experience of encoding protocols in BSPL.  They observe that BSPL is
``more of a core calculus than a usable notation.''  Recent work on
meaning-based languages that compile into BSPL appears to supports the
idea that BSPL represents a core calculus \citep{singh:clouseau:2020}.
\citeauthor{Winikoff+18:HAPN} also report an extensive user study that
evaluated HAPN, AUML, and \review{B:statecharts-b} statecharts for
ease of reading, understanding, and writing specifications.

\citet{ancona:eval:2018} include BSPL, HAPN, and trace expressions in
an evaluation of MAS approaches for support remote patient monitoring.
Although their ten criteria are diverse, a predominant thrust is
tooling, e.g., IDE support, static and runtime verification, testing,
code generation, and so on.  In their evaluation, trace expressions
offer a clear advantage over both HAPN and BSPL in supporting
self-adaptation.  A criterion where both trace expressions and HAPN
shine over BSPL is that they come with IDE support whereas BSPL
doesn't.  In contrast to our and Winikoff {\etal}'s evaluations,
Ancona {\etal}'s evaluation is informed by protocols as use cases.

Together, the three evaluations---the present paper, Winikoff
{\etal}'s, and Ancona {\etal}'s---highlight the variety of criteria
and methods for evaluating protocol languages.  Ultimately, they all
bear upon usability in the larger sense of the term.  Nevertheless, it
is helpful to distinguish \emph{representational} criteria from
\emph{contingent} ones.  A representational criterion is one whose
evaluation for any language relies on primarily on the syntax and
semantics of the language itself.  Each criterion in our evaluation is
representational; we relied on nothing more than the formal
description of the language itself in our analysis.  A contingent
criterion is one whose evaluation is potentially affected to a
significant degree by factors external to the language.  Criteria such
as popularity, ease of use (viewed it as a matter of the familiarity
of the paradigm), and tool support are thus contingent.  In contrast
to our evaluation, both Winikoff {\etal}'s and Ancona {\etal}'s
evaluations have a strongly contingent flavor.

\subsection{Directions}

\review{D:future}With respect to future evaluations of protocol
languages, two directions (with a more contingent flavor) stand out,
especially since they are informed by what would be perceived as broad
strengths of the languages that BSPL outperformed in the current
evaluation.  One, perform a comparative study of protocol languages
for the kinds of properties that can be verified for both protocols
and endpoints.  Scribble and the Trace approaches may do better in
such an evaluation given that verification (both static and runtime)
has been the primary motivation in their development and both benefit
from deep connections with programming language theory.  Two, perform
a comparative study of the programming models supported by the
languages in terms of how they make it easier to write correct
programs.  Scribble, notably, has implementations in several popular
languages such as Java and Python.

A more theoretical direction would be to determine the fragment of a
protocol language that can be encoded in another with the purpose of
formally establishing their relative expressiveness.  We hope that the
evaluation presented in this paper, bearing as it does solely on
representational issues, will help inform such an effort.

% Because a protocol signifies order and asynchrony disorder, there is a
% natural tendency to reconcile them via a combination of significant
% language restrictions (for example, no mixed choice) and operational
% assumptions (for examaple, FIFO), as is done, for example, in
% Scribble, Trace-C, and Trace-F.  As we saw above, by specifying
% causality, BSPL reconciles the two without significant concessions.

\textit{Acknowledgments.}  Raymond Hu provided helpful suggestions
regarding Scribble.  Viviana Mascardi and Angelo Ferrando helped us
understand Trace-F.

Thanks to the anonymous reviewers for helpful comments.  Christie,
Smirnova, and Chopra were supported by EPSRC grant EP/N027965/1
(Turtles). Christie and Singh were partially supported by the National
Science Foundation under grant IIS-1908374.

%Chopra was supported by EPSRC grant EP/N027965/1 (Turtles).  Singh
%thanks IBM and the US Department of Defense for partial support under
%the Science of Security Lablet.

\bibliographystyle{apacite}

% \bibliography{Munindar,Samuel,Amit}

%\clearpage		%\cleardoublepage
%\input{JAIR-responses-mps-v3}

\end{document}